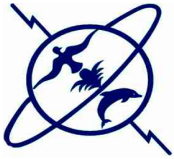

Master Informatique Appliquée

Département Mathématiques et Informatiques

Rapport de Projet de Fin d'Etude

# Infrastructure Logicielle d'un Environnement Hospitalier Intelligent

Youssef GAHI [1], Meryem LAMRANI [1], Mouhcine GUENNOUN [2], and Khalil EL-KHATIB [2]

[1] *Département Math-Info, Faculté des Sciences de Rabat, 4 Avenue Ibn Battouta B.P. 1014 RP, Rabat, Maroc.*

[2] *University of Ontario Institute of Technology, 2000 Simcoe Street North, Oshawa, Ontario, Canada. L1H 7K4.*

Juillet 2008

# Abstract

The impact of new technologies in the field of healthcare has been proven to be satisfactory in improving the quality of care and services for patients. The hospital environment is undoubtedly one of the environments where human or hardware errors can cause harmful damage. It is therefore very useful for the sector to be at the forefront of technology and use the strongest and most efficient means since it is supposed to be an environment of reliability, trust and is legally responsible for ensuring the confidentiality of all patient information. In this project we have proposed and developed a smart hospital enviornement that allows both, detecting patien's fall and send alerts to the suitable parties, and securing patien's stored data through RFID technology.

# Sommaire

# 1 List of Figures

# 2  Introduction

L'impact des nouvelles technologies dans le domaine de la santé a longtemps fait ses preuves et s'avère très satisfaisant quant à l'amélioration de la qualité des soins et des services des patients. L'environnement hospitalier est sans aucun doute l'un des environnements où les erreurs humaines ou matérielles peuvent causer des dommages néfastes. Il est donc d'une grande utilité pour ce secteur d'être à la pointe de la technologie et utiliser les moyens les plus robustes et les plus performants puisqu'il est censé être un milieu de fiabilité, de confiance et est légalement tenu de veiller à la confidentialité de toute information relative aux patients.

La mobilité est d'autre part un facteur indispensable aux professionnels de la santé. Ce secteur étant l'un des premiers à avoir adopté la mobilité comparé aux autres secteurs d'activité. L'utilisation des ordinateurs portables ne cesse d'augmenter notamment grâce à l'intégration des réseaux sans fil. En effet la possibilité pour les médecins et les infirmiers de se déplacer d'un patient à un autre ainsi que la disponibilité immédiate de l'information sur les patients au sein d'un environnement hospitalier et même à l'extérieur de ce dernier sont signes de qualité et d'efficacité des soins prodigués.

# 3  Smart Hospital Environment:

En vieillissant, les êtres humains requièrent beaucoup d'attention. Ils sont en effet facilement victimes de simples incidents telles qu'une chute. Cette dernière est considérée comme étant la première cause de décès accidentel chez les plus de 65 ans. Les causes sont principalement liées à des facteurs propres aux personnes âgées, à l'emploi de certains médicaments, ou à des facteurs environnementaux et les conséquences dépassent la douleur physique puisqu'ils ont aussi un impact psychologique comme la peur de tomber qui est un phénomène courant chez les personnes âgées et qui peut inciter ces derniers à être moins actifs compromettant ainsi leur indépendance.

L'amélioration de la qualité de vie pour les vieilles personnes fait l'objet de plusieurs recherches d'autant plus qu'on prévoit une augmentation de l'espérance de vie moyenne de 6 à 7 ans d'ici 2060. L'un des moyens vers lequel s'orientent ces recherches est la création d'un environnement de maisons intelligentes (plus connu sous le nom de Smart Hospital

Environment). Ce terme a été introduit il y a une dizaine d'années pour faire référence à l'introduction des dispositifs et des services à l'intérieur de la maison.

Sachant le coût exorbitant d'une journée d'hospitalisation à domicile, on réalise mieux l'enjeu socio-économique que représente l'orientation vers une telle solution.

Dans un but visant à promouvoir un style de vie actif aux personnes âgées, améliorer leur autonomie et permettre à un grand nombre de personnes d'accéder au progrès de la médecine tout en maîtrisant le coût des dépenses liées à la santé et à la dépendance des vieux, notre projet s'oriente essentiellement vers la construction d'un système pour un environnement hospitalier intelligent qui permet d'une part la détection des chutes sans le besoin d'une assistance humaine réduisant ainsi les coûts liés aux soins à domicile. D'autre part, le contrôle d'accès aux données personnelles des patients leur garantissant ainsi une totale confidentialité.

# 4  Cas d'Utilisation:

Agé de 70 ans, Monsieur J. ne renonce pas à son indépendance en décidant de continuer à vivre tout seul chez lui sans aucune assistance. Son fils occupé par son travail ne peut pas toujours lui rendre visite aussi souvent qu'il le souhaiterait. Face à la cherté des soins médicaux à domicile, il opte donc pour un dispositif d'accéléromètre intégrant un « heart Beat Monitor » qui envoie continuellement à l'ordinateur personnel de Monsieur J. les informations sur l'état du cœur de ce dernier ainsi que le suivi de ses mouvements.

En installant le logiciel du dispositif, monsieur J. accepte de se conformer aux conditions d'utilisation qui donne le droit à l'ordinateur de composer le numéro du service des urgences sans attendre son consentement.

Peu après l'acquisition de cet appareil, Monsieur J. subi une chute alors qu'il était seul chez lui. Il resta suffisamment allonger au sol pour que le cas soit considéré urgent. L'ordinateur détecte la chute grâce à un algorithme qui calcule les données de l'accéléromètre et effectue l'appel aux urgences qui arrivent sur place peu de temps après.

Néanmoins, le traitement efficace de monsieur J. requière la connaissance par le médecin de son état de santé et donc de son dossier médical. Ce dernier dispose donc d'un badge RFID qui permet de l'authentifier comme étant un profil valide. L'ordinateur se déverrouille et se connecte à la base de données qui lui donne l'accès aux informations personnelles du patient.

Une fois que le médecin termine avec la machine, cette dernière - en ne détectant plus la proximité de la carte RFID - se verrouille à nouveau et perd la connexion avec la base de données rendant de ce fait l'accès non permis à une tierce personne.

## 5   Architecture de l'Application:

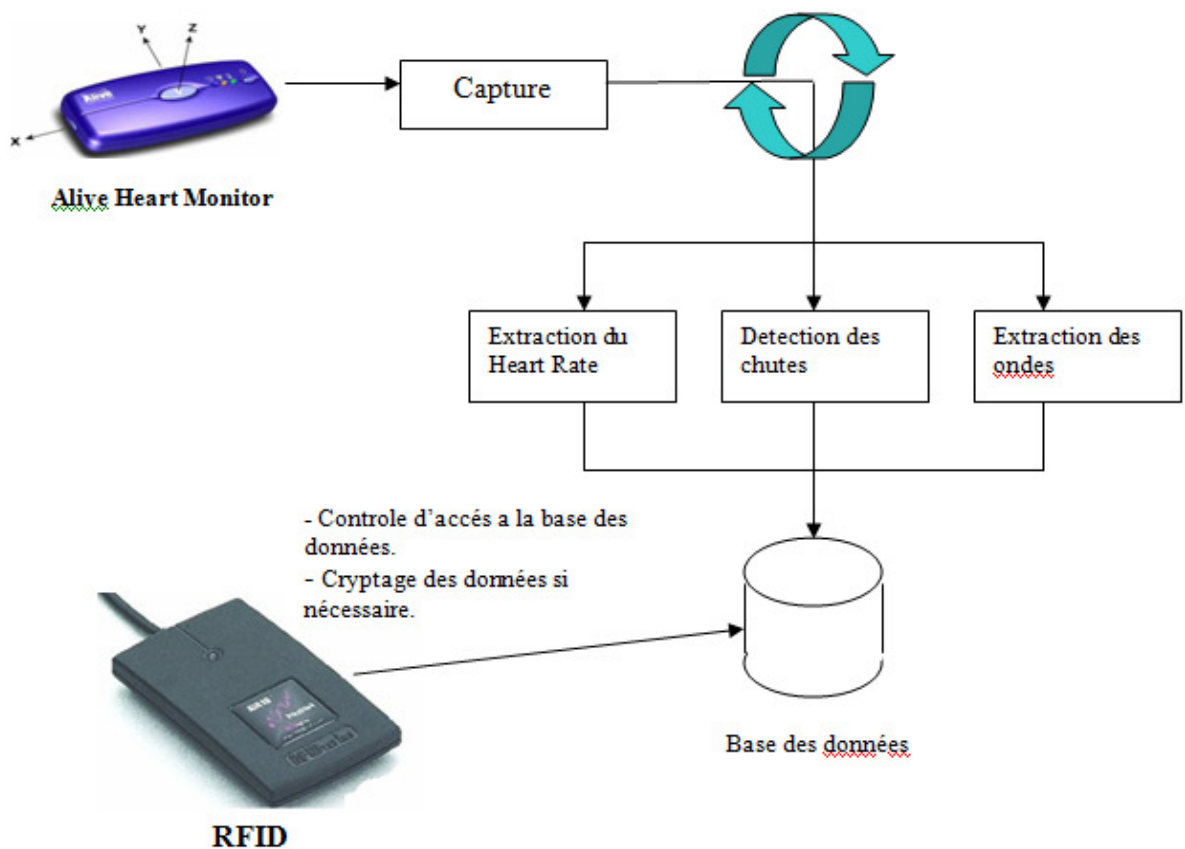

Figure 5-1: Architecture de l'application.

# 6 Extraction des données ECG et ACC
## 6.1 Alive Heart Monitor
### 6.1.1 Fonctionnement du Alive Heart Monitor

Alive Heart Monitor est un appareil qui renseigne sur l'activité cardiaque de la personne l'utilisant ainsi que sur le suivi de ses mouvements selon les trois axes (X, Y, Z). Pour cela, il dispose d'une connexion Bluetooth. Cette dernière lui permet de communiquer avec n'importe quelle machine qui supporte cette technologie (Pc, téléphone mobile,…).

Dans le cadre de ce projet, la communication se fait avec une clé Bluetooth branchée au port série d'un ordinateur.

### 6.1.2 Les données envoyées

Une fois la connexion établie, le moniteur envoie au PC des paquets contenants les informations suivantes:

#### 6.1.2.1 ECG

Les données ECG sont des informations qui étudient le fonctionnement du cœur en mesurant son activité électrique. A partir de ces informations, on peut déterminer si l'activité électrique est normale, irrégulière ou rapide.

Le moniteur envoie ces ECG avec un débit de 300 (échantillons/seconde) sous forme d 'entiers compris entre 0 et 255. Ce qui est équivaut à -2.66 et 2.66 mV.

#### 6.1.2.2 Accélérations des 3 Axes

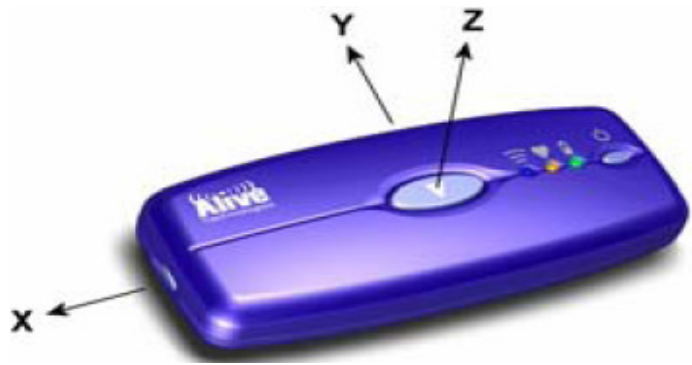

Figure 6-1: Alive Heart Monitor.

Le moniteur dispose d'un accéléromètre qui mesure l'accélération du moniteur suivant les 3 Axes X,Y et Z, ces valeurs sont des entiers compris entre 0 et 255, équivaut à -2.7g et + 2.7g.

### 6.1.2.3 Format des données

Les informations envoyées par le moniteur peuvent être récupérées, soit en temps réel, via la connexion SPP (Serial Port Profile), ou alors à partir d'un fichier de stockage. Les paquets envoyés par le moniteur respectent le format suivant:

**Packet Structure**

| Packet Header 6 Bytes | ECG Header 5 Bytes | ECG Data n Bytes | Acc Header 5 Bytes | Acc Data n Bytes | Checksum 1 Byte |
|---|---|---|---|---|---|

Figure 6-2: Format du paquet.

La taille du paquet est calculée selon la relation suivante: (17 + m + n) octets. m et n étant respectivement la taille en octets des informations ECG et Accélération. Ces deux entiers sont déterminés dans les ECG Header et ACC Header.

**Packet Header**

| Byte | Value | Description |
|---|---|---|
| 1 - 2 | Sync Bytes | Each packet starts with 1$^{st}$ byte 0x00 2$^{nd}$ byte 0xFE |
| 3 | Battery Level (%) | 0-200 (200 = 100%) |
| 4 - 5 | Sequence number Bits 0 – 11 | A 12bit number that is incremented for each successive packet. |
|  | Status Bits 12 – 15 | Bit 12 – Event Bit 13 – Reserved Bit 14 – Reserved Bit 15 – Reserved |
| 6 | Number of data blocks |  |

Figure 6-3: Format de l'entête du paquet.

La taille de l'entête du paquet est de 6 octets. Les deux premiers octets contiennent la valeur hexadécimale 0x00FE qui désigne le début d'un nouveau paquet. Le troisième octet contient le niveau de la batterie du moniteur. Les 12 bits qui suivent représentent le numéro de séquence du paquet courant. Et enfin le dernier octet détermine le nombre de blocs de données.

## ECG Header

| Byte | Value | Description |
|---|---|---|
| 1 | ID | 0xAA = ECG |
| 2 - 3 | Length | Length of ECG header and ECG data in bytes |
| 4 | Data Format | 0x01  Data Type:    8bit unsigned<br>Sampling rate: 150 samples/s<br>Range:         +- 2.66mV<br><br>0x02  Data Type:    8bit unsigned<br>Sampling rate: 300 samples/s<br>Range:         +- 2.66mV |
| 5 | Reserved | Not used |

Figure 6-4: Format de l'entête du paquet ECG

L'entête des données ECG est composé de 5 octets. Le premier contient l'identificateur du paquet, le $2^{ème}$ et le $3^{ème}$ déterminent la taille de l'entête ainsi que les données des informations ECG, et le $4^{ème}$ détermine le nombre d'échantillons par seconde.

## Acc Header (3 Axis)

| Byte | Value | Description |
|---|---|---|
| 1 | ID | 0x56 = 3 Axis Accelerometer |
| 2 - 3 | Length | Length of Acc header and Acc data in bytes |
| 4 | Data Format | 0x00  Data Type:         8 bit unsigned<br>Sampling rate:      75 samples/s<br>Range:              +- 2.7g<br>Interleaved samples: $X_1Y_1Z_1X_2Y_2Z_2X_3Y_3Z_3...$ |
| 5 | Reserved | Not used |

Figure 6-5: Entête du paquet ACC.

L'entête des données ACC est composée également de 5 octets, le premier contient l'identificateur du paquet, le $2^{ème}$ et le $3^{ème}$ déterminent la taille de l''entête ainsi que les données des informations ACC et le $4^{ème}$ détermine le nombre d'échantillons par seconde.

Les données sont stockées dans les fichiers (ATS) sous la structure suivante :

### Structure

| File Header | Main Header |
|---|---|
| | Channel 1 Description |
| | Channel 2 Description |
| | Channel 3 Description |
| | ... |
| | Channel n Description |
| 1st Data Block | Channel 1 Data Packet |
| | Channel 2 Data Packet |
| | Channel 3 Data Packet |
| | ... |
| | Channel n Data Packet |
| 2nd Data Block | Channel 1 Data Packet |
| | Channel 2 Data Packet |
| | Channel 3 Data Packet |
| | ... |
| | Channel n Data Packet |
| 3rd Data Block | Channel 1 Data Packet |
| | Channel 2 Data Packet |
| | Channel 3 Data Packet |
| | ... |
| | Channel n Data Packet |
| ... | ... |

Figure 6-6: Structure du fichier ATS.

### Main Header

| Byte | Value | Description |
|---|---|---|
| 1-5 | ID | ATSF<br>Bytes: 0x41, 0x54, 0x53, 0x46, 0x00 |
| 6-7 | Length of Header | Length of File Header in Bytes |
| 8 | Channels | Number of Data Channels |
| 9-12 | Num of Data Blocks | Total Number of Data Block<br>0 if not known |
| 13-14 | Length of a Data Block | Number of bytes per Data Block |
| 15-18 | Date | Byte 1-2    Binary: Year (eg 2005)<br>Byte 3      Binary: Month (1-12)<br>Byte 4      Binary: Day (1-31) |
| 19-21 | Time | Byte 1      Binary: Hour (0 - 23)<br>Byte 2      Binary: Minute (0-59)<br>Byte 3      Binary: Second (0-59) |
| 22-128 | Reserved | |

Figure 6-7: Structure de l'entête principal.

## Channel Description

| Byte | Value | Description |
|---|---|---|
| 1 | Data Type | 0x11 = Status |
| | | 0xAA = ECG |
| | | 0x55 = 2 Axis Accelerometer |
| | | 0x56 = 3 Axis Accelerometer |
| 2 | Data Format | Specifies the format of the data |
| 3-4 | Length | Number of bytes per channel Data Packet |
| 5-32 | Reserved | |

Figure 6-8: Format de la descritpion canal.

## Data Channels

| Data Type | | Data Format | Description | |
|---|---|---|---|---|
| 0x11 | Status | 0x00 | Byte 1 | Status Bits |
| | | | | Bit 0 - Reserved |
| | | | | Bit 1 - Reserved |
| | | | | Bit 2 - Reserved |
| | | | | Bit 3 - Reserved |
| | | | | Bit 4 - Reserved |
| | | | | Bit 5 - Reserved |
| | | | | Bit 6 - Reserved |
| | | | | Bit 7 – Button Event |
| | | | Byte 2 | Battery voltage (200−100%) |
| 0xAA | ECG | 0x01 | Data Type: | 8bit unsigned |
| | | | Sampling rate: | 150 Hz |
| | | | Range: | +- 2.66mV |
| | | 0x02 | Data Type: | 8bit unsigned |
| | | | Sampling rate: | 300 Hz |
| | | | Range: | +- 2.66mV |
| 0x55 | 2 Axis Accelerometer | 0x00 | Data Type: | 8bit unsigned |
| | | | Sampling rate: | 75 Hz |
| | | | Range: | 2g |
| | | | Interleaved samples: | $X_1Y_1X_2Y_2...$ |
| 0x56 | 3 Axis Accelerometer | 0x00 | Data Type: | 8bit unsigned |
| | | | Sampling rate: | 75 Hz |
| | | | Range: | +-2.7g |
| | | | Interleaved samples: | $X_1Y_1Z_1X_2Y_2Z_2X_3Y_3Z_3...$ |

Figure 6-9: Format des données du canal.

## 6.2 Plateforme logicielle pour l'extraction des données du moniteur

### 6.2.1 Service RFCOMM et le profil SPP

Un profil correspond à une spécification fonctionnelle d'un usage particulier. Il correspond notamment à différents types de périphériques. Les profils ont pour but d'assurer une interopérabilité entre tous les appareils Bluetooth. Ils définissent la manière d'implémenter un usage défini les protocoles spécifiques à utiliser.les contraintes et les intervalles de valeurs de ces protocoles.

RFCOMM est un service basé sur les spécifications RS-232, qui émule des liaisons séries. Il peut notamment servir à faire passer une communication IP par Bluetooth.

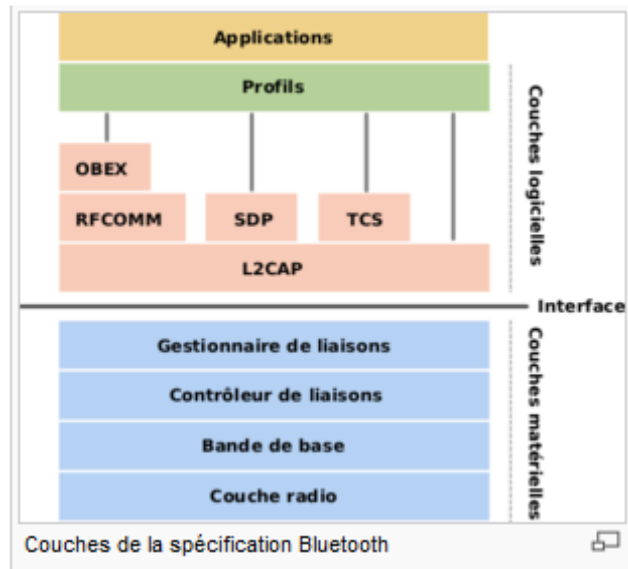

Figure 6-10: La pile bluetooth.

## 6.2.2 API BlueCove

Afin de capturer les données envoyées par le moniteur, nous avons utilisé l'API BlueCove qui est une interface logicielle avec la pile Bluetooth.

BlueCove est un API JSR82 développé par une équipe Intel afin d'assurer une communication entre un appareil mobile et un PC disposant d'une connexion Bluetooth. BlueCove définit donc des classes et des méthodes simples comme connect, read et write, afin d'établir une communication fiable. L'utilisateur peut donc se contenter d'utiliser ses méthodes sans avoir besoin de connaître la structure de la pile Bluetooth. Pour se connecter au matériel il faut connaitre le type de la connexion (RFCOMM, L2CAP...), le numéro de port où la clé est branchée et l'adresse physique de la clé. Ces informations seront regroupées dans un lien de connexion tel : **btspp://Bluetooth_Adresse_physique : port**

Pour capturer les données envoyées par le Bluetooth, nous avons développé une application java sous éclipse, qui utilise l'API BlueCove 2.0, ajouté à un projet éclipse comme un jar externe, et ce selon les étapes suivantes :

Project → Proprietes → Java Build Path→Add External Jar

Ajouter: BlueCove 2.0.zip

### 6.2.3 Visual Studio 8

Le choix de Visual Studio comme plateforme est dû au besoin de réutiliser un programme développé en C++ et qui permet de détecter des périodes à partir des valeurs ECG. Développé par la compagnie Eplimited, Ce programme analyse de valeurs ECG, construit un signal périodique détecte les sommets du signal. Le nombre de pic par minutes sera donc la valeur de heart Rate de la personne concernée.

# 7 Détection des Chutes

La deuxième partie de notre application repose sur la détection des chutes. Comme nous avons déjà cité le moniteur dispose d'un accéléromètre qui mesure l'accélération de chaque axe du moniteur.

## 7.1 Pourquoi une détection des chutes ?

Le taux de vieillissement au Canada ne cesse de croître. Chaque année on recense plus de 11 millions de personnes âgées ayant été victimes de chutes. Selon les statistiques 60% de ces chutes se produisent à leur domicile alors qu'elles ne bénéficient d'aucune assistance, d'où le grand intérêt de développer une application qui surveille les mouvements des patients pour une détection efficace et rapide des chutes.

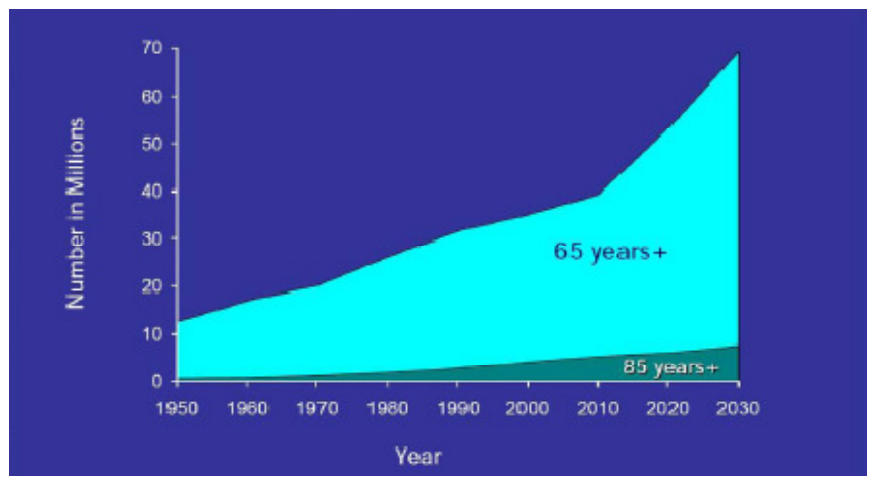

Figure 7-1: Taux de vieillissement au Canada.

## 7.2 L'Accéléromètre

Un **accéléromètre** est un capteur fixé à un mobile. Il permet de mesurer l'accélération de ce dernier. Cette accélération est exprimée en m/s².

Le principe de tous les accéléromètres est basé sur la loi fondamentale de la dynamique **F = M. a** (F: force, M: masse, a: accélération notée aussi gamma). Plus précisément, il consiste en l'égalité de la force d'inertie, de la masse sismique du capteur et d'une force de rappel appliquée à cette masse.

## 7.3 La détection d'une chute à partir des valeurs d'accélérations

Une chute se définit tout simplement comme étant un mouvement accéléré vers le sol. Les valeurs d'accélérations envoyées par le moniteur suivant chaque axe (X, Y et Z), peuvent être interprétées afin de prouver qu'une chute à eu lieu. Il existe plusieurs manières pour utiliser la variation des valeurs des accélérations afin de détecter une chute. On peut utiliser par exemple les changements dans les angles du repère suivant les coordonnées sphériques.

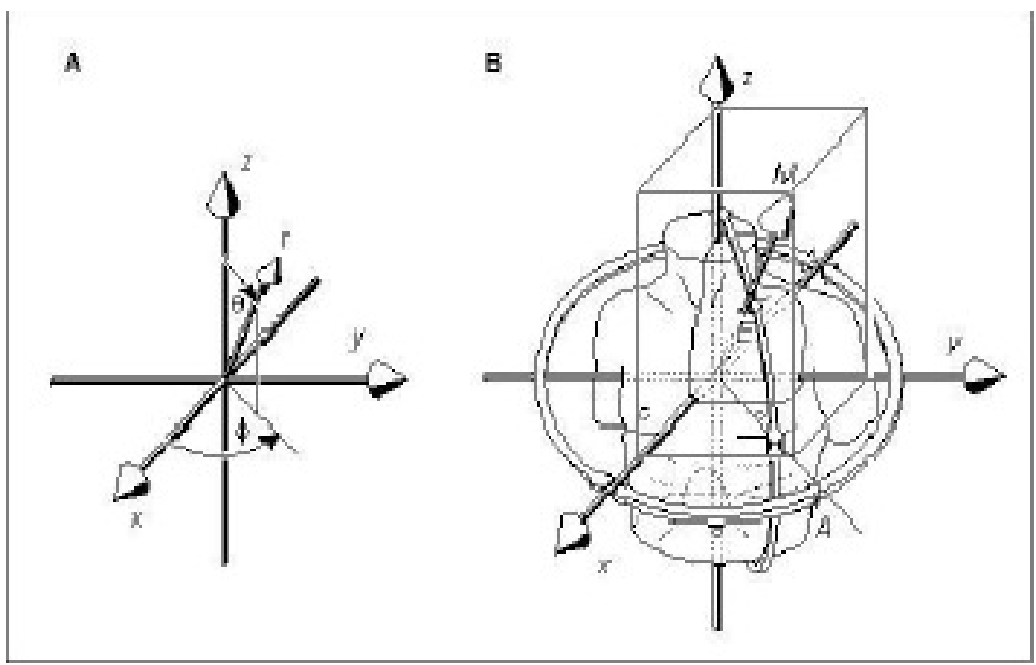

Figure 7-2: Principe de surveillance des mouvements.

D'autre part, on peut tester après une variation dans les valeurs des accélérations si le moniteur est horizontal au sol. Si ce dernier maintient la même position pendant une certaine durée (8 secondes par exemple), on pourra confirmer que le patient a eu une chute.

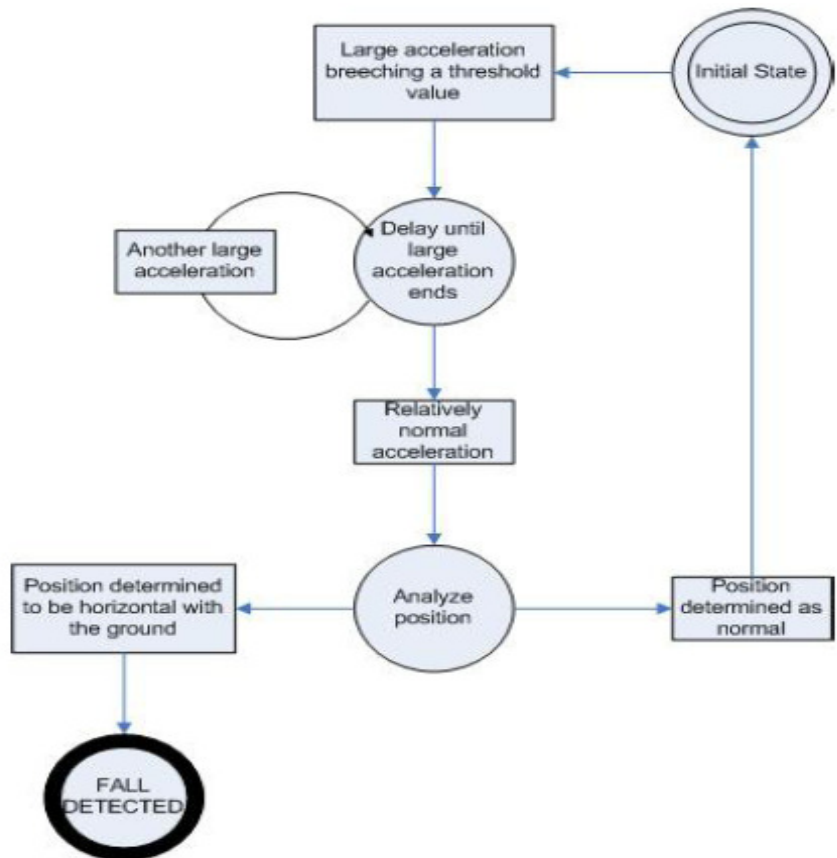
Figure 7-3: Algorithme de détection des chutes.

# 8 Identification par ECG

Le signal généré à partir des valeurs d'électrocardiogramme peut être utilisé pour identifier les personnes. D'après des études et des statistiques la période d'un signal est contenue dans des petits intervalles de temps qui sont uniques pour chaque personne.

## 8.1 Les différents intervalles d'un signal ECG

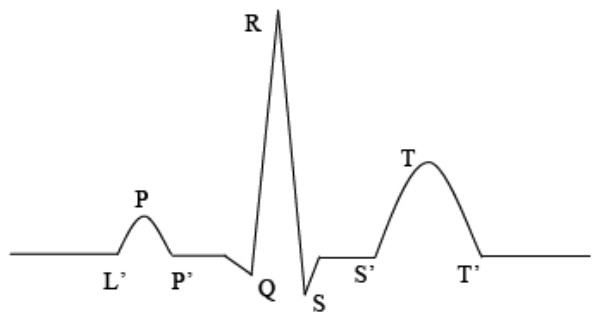
Figure 8-1: Signal ECG.

Comme nous avons déjà expliqué dans la première section, le signal d'électrocardiogramme (ECG) détermine l'activité du cœur. Chaque période du signal pour un battement normal présente 3 petits intervalles (P, QRS, T). L'onde P a la fréquence la plus basse dans une période du signal, elle est comprise entre 10 et 15 Hz, et elle dure moins de 120 ms. Le QRS a normalement la plus grande fréquence (entre 10 et 40 Hz pour un battement de cœur normal) et elle dure entre 70 - 110 ms. Les durées de chaque intervalle ainsi que la hauteur des ondes sont distincts d'une personne à une autre, nous nous sommes donc basés sur ces points pour implémenter notre approche.

Les données ECG passent tout d'abord par un classificateur qui détecte les périodes bruitées. Puis on traite ces signaux pour extraire les durées de chaque onde.

## 8.2 La distance de Mahalanobis pour la discrimination

Notre approche a pour but l'identification des personnes et le stockage dans une base de données de plusieurs profils concernant différentes personnes (moyennes et variance pour chaque paramètre du signal). Pour vérifier un profil parmi les profils existants on calcule sa moyenne, sa variance, puis on calcule sa distance de Mahalanobis par rapport à tous les profils existants. La distance minimale serait donc celle du profil correspondant.

La distance de Mahalanobis est une mesure de distance introduite par P. C. Mahalanobis en 1936. Elle est basée sur la corrélation entre des variables par lesquelles différents modèles peuvent être identifiés et analysés. C'est une manière utile de déterminer la *similarité* entre une série de données connues et inconnues. Elle diffère de la distance euclidienne par le fait qu'elle prend en compte la corrélation de la série de données. Ainsi, à la différence de la distance euclidienne où toutes les composantes des vecteurs sont traitées de la même façon, la distance de Mahalanobis accorde un poids moins important aux composantes les plus bruitées (en supposant que chaque composante est une variable aléatoire de type gaussien).

La distance de Mahalanobis est souvent utilisée pour la détection des données aberrantes dans un jeu de données, ou alors pour déterminer la cohérence de données fournies par un capteur. Exemple : cette distance est calculée entre les données reçues et celles prédites par un modèle.

Pratiquement, la distance de Mahalanobis d'une série de valeurs de moyenne $\mu = (\mu_1, \mu_2, \mu_3, \ldots, \mu_p)$ et possédant une matrice de covariance Σ pour un vecteur à plusieurs variables $x = (x_1, x_2, x_3, \ldots, x_p)$ est définie comme suit:

$$D_M(x) = \sqrt{(x-\mu)^T \Sigma^{-1} (x-\mu)}.$$

## 8.3 Amélioration de l'algorithme de Kyoso

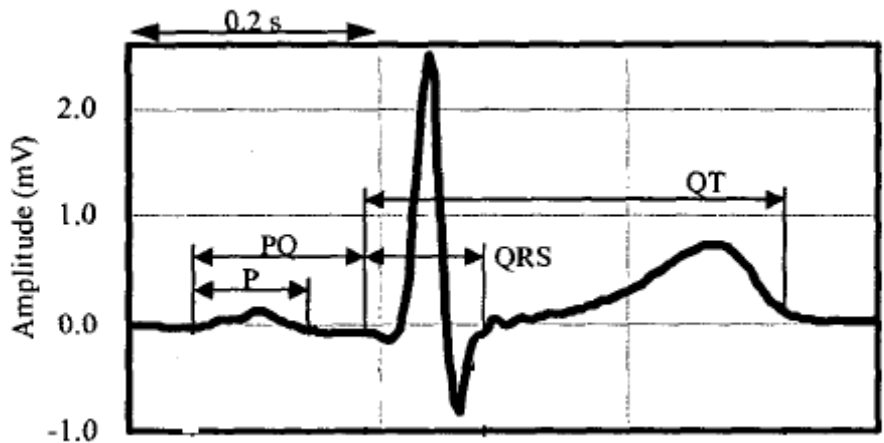

Figure 8-2: Les différentes périodes d'un signal ECG.

L'algorithme de Kyoso se base sur les 4 intervalles (P, PQ, QRS, QT) pour identifier chaque personne, et il utilise les combinaisons entre ces 4 paramètres pour calculer la distance de Mahalanobis

TABLE IV
RESULTS OF DISCRIMINANT ANALYSIS

| Data ID | Accuracy [%] | | | | | |
|---|---|---|---|---|---|---|
| | P-QRS | P-PQ | P-QT | QRS-PQ | QRS-QT | PQ-QT |
| A_2 | 80.8 | 31.3 | 48.7 | 99.1 | 99.1 | 48.2 |
| B_2 | 73.2 | 38.5 | 4.1 | 89.0 | 90.4 | 46.4 |
| C_2 | 84.0 | 90.1 | 81.0 | 95.0 | 76.1 | 78.1 |
| D_2 | 60.3 | 56.3 | 99.4 | 44.8 | 99.4 | 99.4 |
| E_2 | 82.9 | 17.5 | 45.8 | 93.8 | 90.2 | 47.6 |
| F_2 | 63.6 | 7.5 | 86.6 | 67.4 | 99.2 | 87.0 |
| G_2 | 64.3 | 94.9 | 38.4 | 98.4 | 96.5 | 96.9 |
| H_2 | 80.4 | 84.4 | 29.5 | 79.0 | 97.3 | 24.1 |
| I_2 | 99.6 | 88.5 | 83.2 | 99.6 | 99.6 | 84.5 |

Figure 8-3: Résultats des tests de Kyoso.

Le tableau en haut représente les résultats obtenus pour 9 personnes à partir de l'algorithme de Kyoso. Nous remarquons bien qu'ils ne sont pas tout a fait parfaits.

Notre approche pour améliorer l'algorithme de Kyoso est d'utiliser plus d'intervalles, et plus de combinaisons. Voici ci-dessous les différents intervalles sur lesquels nous avons travaillé:

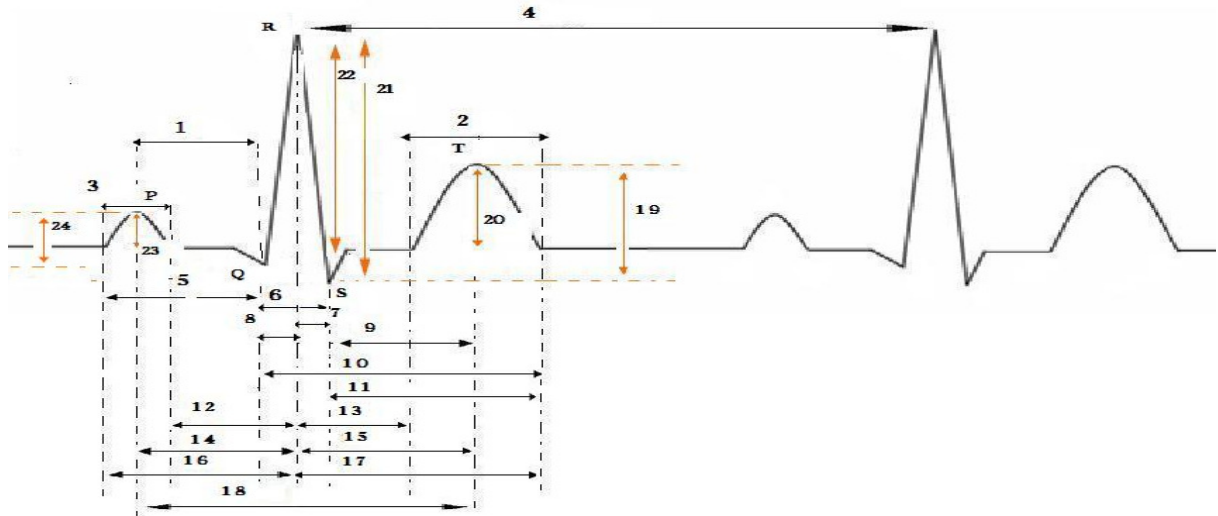

Figure 8-4: 24 caractéristiques pour une période ECG.

# 9 Conception du Système

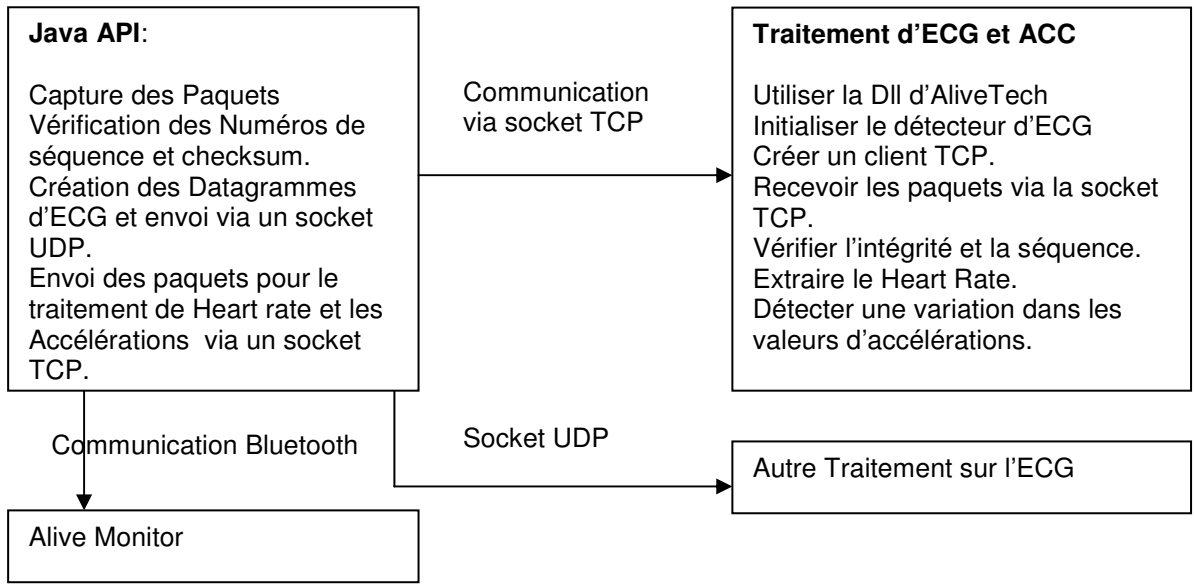

## 9.1 Capture des données

Le composant de capture des données contient 7 classes et une interface dont voici un schéma qui montre la hiérarchie:

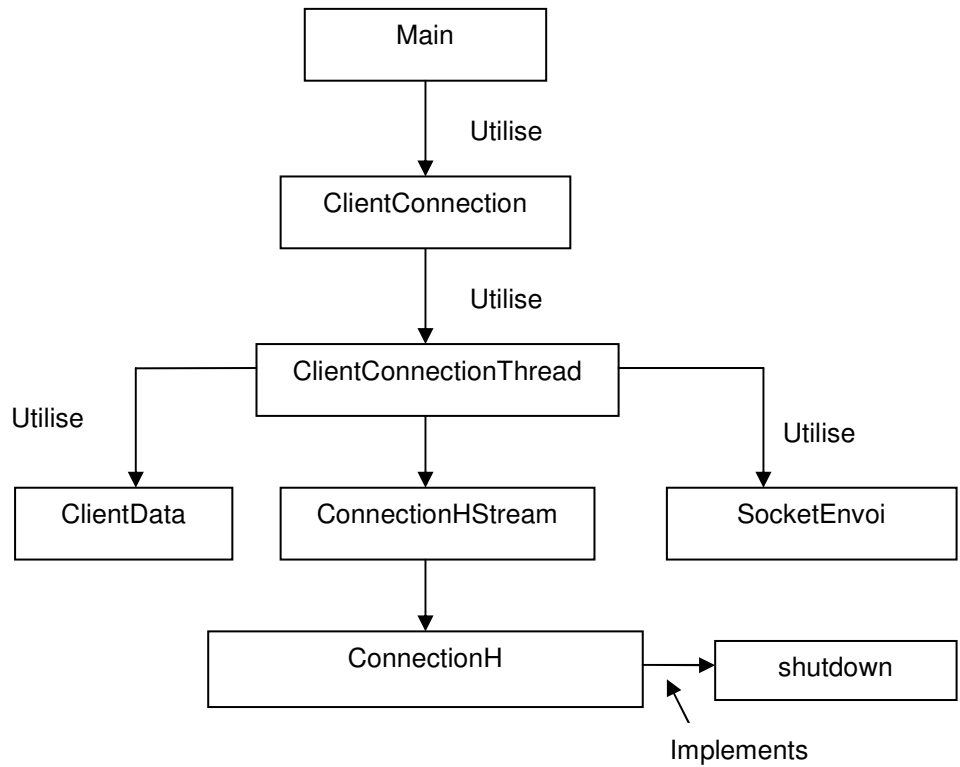

- **Shutdown :** Est une interface qui contient une méthode « Shutdown » pour la fermeture du Thread et lance une interruption en cas d'erreur.

- **ConnectionH :** Classe permettant une amélioration dans des versions ultérieures, elle permet de faire des Connexions en parallèles.

- **ConnectionHStream :** Permet de créer deux Flux de Connexion, l'un pour les Entrées, et l'autre pour les Sorties, ainsi qu'une méthode pour la fermeture de ces flux.

- **SocketEnvoi :** Classe qui crée un Socket TCP. Cette dernière envoie au port 9999 les paquets reçus vers l'autre sous application de C++.

- **ClientData :** Classe qui crée un Socket UDP.

- **ClientConnectionThread :** Classe qui crée une connexion avec le Monitor en utilisant la classe : Connector et qui lance un Thread qui aura pour rôle la lecture des données venant via la connexion Bluetooth à travers la fonction read de la classe, Connexion.

Cette classe utilise la classe SocketEnvoi pour lancer le serveur d'envoi, et une fois qu'un client se connecte au serveur le Thread commence à lire les données. Par la suite, après chaque 143 octets, le thread utilise la méthode Send de la classe SocketEnvoi pour envoyer ces 143 octets reçus, et ainsi de suite. Une fois que le thread lit l'entier -1 comme valeur d'un octet, cela veut dire que la connexion est terminée, donc le thread termine son travail en tuant les flux de connexions.

- **ClientConnection :** Classe qui permet de manipuler la classe « ClientConnectionThread » en utilisant deux méthodes :

  **Connect** : qui permet de lancer un thread.

  **Shutdown** : qui permet de tuer le thread.

- **Main :** Classe mère de cette sous application, elle lance une fenêtre avec deux boutons :

  - **Bouton Connect**: qui permet de lancer un Client de type : « ClientConnection»
  - **Bouton Disconnect**: qui termine ce client.

## 9.2 Extraction de Heart Rate a partir des ECG

Les ECG reçus par l'application java ne représentent que l'état électrique du cœur, il faut donc bien évidement une interprétation pour extraire les valeurs du heart rate. Le choix de la plateforme Visual Studio est dû au besoin de réutiliser des composantes implémentées en C++ permettant de générer des périodes et des fréquences à partir des valeurs entières des ECG. Nous avons donc créés une socket qui récupère les données capturées par le programme java. On revérifie par la suite le checksum et le numéro de séquence pour chaque paquet reçu. Enfin on fait passer ces entiers à l'application C++ qui retourne le heart rate.

## 9.3 Detection des chutes

Le moniteur Alive Heart envoie des accélérations suivant chaque Axe de son repère qui a comme point d'origine son centre d'inertie. Ces accélérations sont des valeurs numériques comprises entre **0 et 255** et qui correspondent à **-2.7g et 2.7g.** Les valeurs entre 0 et 127 sont des valeurs négatives c.-à-d. une accélération dans le sens contraire de l'orientation d'axe correspondant.

**Remarque**: Dans le cas où le monitor est en parallèle au sol , les valeurs initiales sont X = 127 , Y = 127 ,    Z = 175 ou Z = 80 , les deux valeurs de X et Y sont 0 mais la valeur de Z signifie qu'il y a toujours une force de gravitation.

**Approche**:

Nous avons constaté que lorsque le monitor fait une accélération, il entre dans deux cycles. Le cycle de mouvement et le cycle de stabilisation : lorsqu'on soulève un monitor qui était à l'état initiale (l'axe Z est perpendiculaire), on remarque que les valeurs de Z vont monter pendant 1/3 secondes et après ils vont descendre jusqu'à la stabilisation. D'où la base avec laquelle nous avons travaillé pour détecter une chute c.-à-d. lorsqu'on lit les informations sur l'accélération des axes en détectant une variation, on commence à voir les 23 valeurs qui suivent. On compare le max ou le min de ces valeurs **( cela dépend de l'orientation de l'axe, si l'axe est au sens contraire de l'orientation pendant une chute les valeurs vont monter et s'il est au bon sens pendant une chute les valeurs vont descendre )** avec la dernière valeur avant la variation , et si on remarque que la différence dépasse un certain seuil et aussi la moyenne des 23 échantillons est plus grande ou plus petite de la dernière valeur avant la variation, on déclare qu'une chute a eu lieu.

Pendant une chute le mouvement ou bien l'accélération du monitor est celle de l'axe le plus perpendiculaire, donc on se base essentiellement pour détecter une chute sur l'accélération de l'axe le plus perpendiculaire (le plus proche de 175 ou 80).

**Résultat**:

Le programme détecte + ou – 90 % des cas de tests.

## 9.4  Système d'Identification par ECG

Après avoir effectué la capture des données et leurs envois à travers un socket vers le système de traitement, les données ECG arrivent en phase de filtrage. Le signal ECG est très souvent bruité, c'est la raison pour laquelle on utilise des filtres afin d'ignorer les ondes défectueuses.

La compagnie Physionet met en disposition des filtres en ligne, à l'adresse : www.physionet.org. Afin d'utiliser ces filtres, le format des ECG doit correspondre au Physionet ECG format. On peut transformer les données ECG en format compatible grâce au programme exécutable wrsamp.exe, qui utilise les 7 dll: « cygcrypto-0.9.8.dll », « cygcurl-4.dll », « cygssh2-1.dll », « cygssl-0.9.8.dll », « cygwfdb-10-4.dll », « cygwin1.dll », « cygz.dll».

Ces dll sont obtenues en installant le WFDB et CYGURL.

La transformation se fait grâce aux commandes suivantes :

**wrsamp -F** 300 **–G** 50 **–i** in.txt **–o** out

-**F**: la fréquence.

-**G**: l'unité par mV.

-**i**: le fichier d'entrée en format texte.

-**o**: les noms des fichiers de sortie en deux formats (out.dat et out.hea).

Le fichier .dat serait l'entrée du filtre utilisé. A partir de résultats ainsi obtenus, on extrait les durées de chaque intervalle, ainsi que l'auteur de chaque onde et la valeur ECG associée. Puis on stocke dans notre base de données la moyenne et la variance pour chaque intervalle parmi les 24 intervalles utilisés dans le système d'enregistrement. Dans la phase de vérification on compare les valeurs récupérées avec les profils stockés dans notre base. En utilisant la distance de Mahalanobis, on prend la décision à qui appartiennent les informations traitées.

## 9.5 Installation du Kit Bluetooth

Pour lire les informations venant directement de la clé Bluetooth, il faut installer tout d'abord le driver Bluetooth qui peut être téléchargé à travers ce lien : http://direct.motorola.com/Hellomoto/NSS/PC850BTSW/

Après le Redémarrage de la machine, il faut tout d'abord lancer la clé Bluetooth en cliquant à travers le bouton droit de la souris sur la petite icone qui représente le programme exécutable du Bluetooth, et qui se trouve avec les programmes en cours d'exécution.

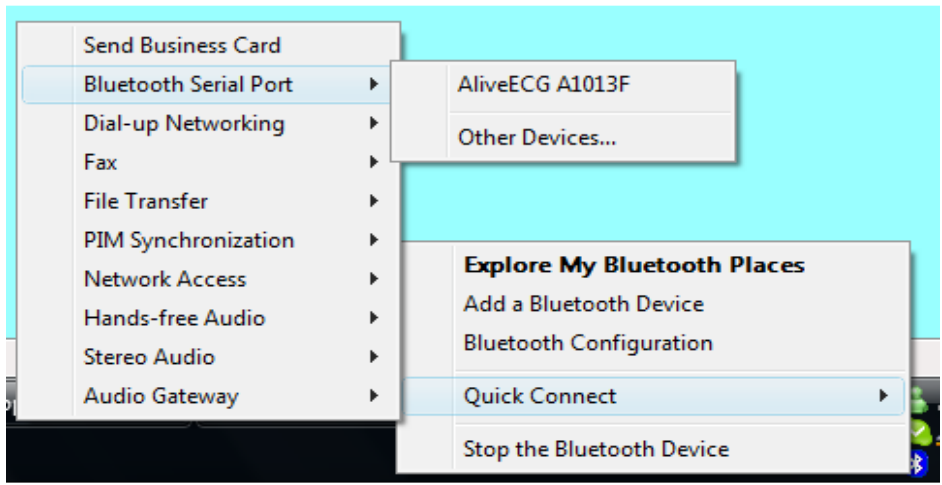

Figure 9-1: 24 connexion au Moniteur.

Après le lancement de la clé Bluetooth, il est nécessaire de configurer la connexion avec le moniteur, en suivant les étapes décrites sur l'image en haut. Il est préférable dans les étapes d'ajout du matériel de désactiver la connexion avec un mot de passe pour faciliter la connexion au périphérique.

On note que le numéro du port de connexion et l'adresse physique du matériel qui figurent dans le sous menu « Bluetooth Configuration ».

## 9.6 Installation de BlueCove

BlueCove 2.0.2 peut être téléchargé à partir du lien suivant : http://sourceforge.net/project/showfiles.php?group_id=114020

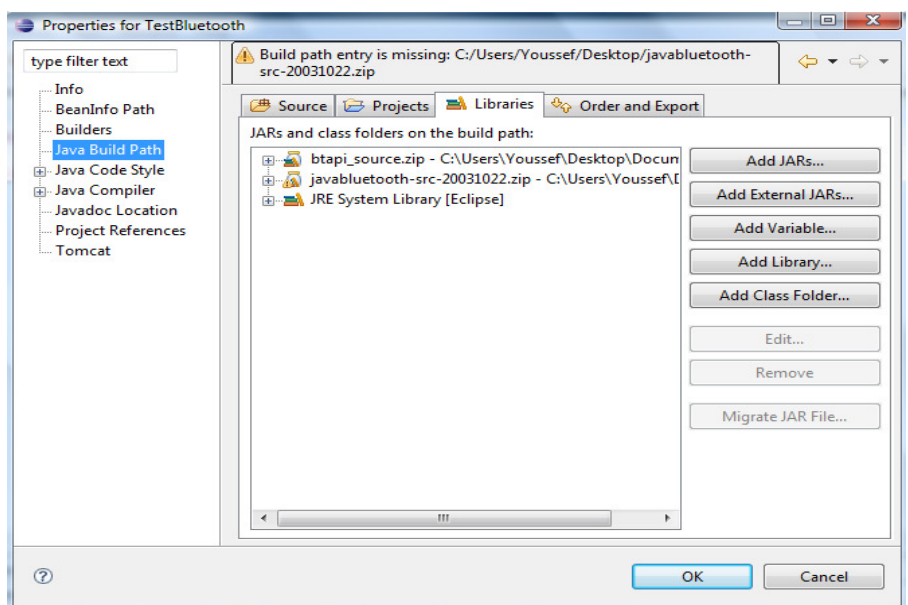

Figure 9-2: Ajout d'un API à un projet Eclipse.

Après le téléchargement de la source blueCove2.0.2.zip on peut l'ajouter à un projet Eclipse en cliquant sur le menu Projet puis Propriétés puis java Build Path, après on peut grâce à l'option Add external Jars ajouter la source au projet.

# 10 Badges RFID

Le personnel d'un hôpital, d'une banque ou d'une entreprise quelconque dispose de nos jours de badges qu'ils présentent à des lecteurs fixés aux murs leur permettant de s'identifier afin d'accéder aux locaux. L'idée est donc venu d'utiliser le même concept avec le pc pour sécuriser l'accès aux données.

Avec la multitude de badges et de lecteurs disponibles aujourd'hui sur le marché, la sélection des fabricants n'est pas toujours évidente. Certains construisent uniquement des composants matériels d'autres construisent tout, y compris un ensemble d'outils logiciels pour transmettre et partager les données RFID.

Le lecteur pcProx USB provient de la compagnie RFIDeas qui se consacre aux solutions relatives à l'Identification et l'Accès. Cette dernière a pour but d'étendre l'utilisation de badges, initialement dédiés à l'accès aux bâtiments, à un grand nombre d'applications. Ces applications incluent le log-on, l'identification des employés, l'accès au réseau local… tout en utilisant l'unique badge de départ.

Le lecteur est une solution plug and play compatible avec plus de 200 millions de badges existants. Ce dernier est muni d'une mémoire flash lui permettant de le configurer rapidement. Il ne nécessite l'installation d'aucun logiciel pour son fonctionnement, néanmoins une application utile à sa configuration est offerte sur le site de son fabricant.

## 10.1 Format de programmation des cartes de proximité HID

Le format décrit la signification d'un nombre ou comment ce dernier doit être utilisé. Un nombre de bits n'indique pas forcément le format sauf pour le standard 26 bits, ce qui signifie qu'une carte d'un nombre de bits spécifique peut avoir une centaine de formats différents. On dispose de deux formats :

### 10.1.1 Le format 26 bits

C'est le format le plus utilisé. Toute personne ou compagnie peut se le procurer. Le nombre total de cartes qu'offre ce format est 16, 711, 425. Etant un chiffre relativement limité, il existe un risque de duplication d'autant plus qu'aucun contrôle ne garantit l'unicité des cartes ni la limitation des commandes. Leur seul avantage réside dans leur simplicité ainsi que leur compatibilité avec les systèmes de contrôle d'accès existants.

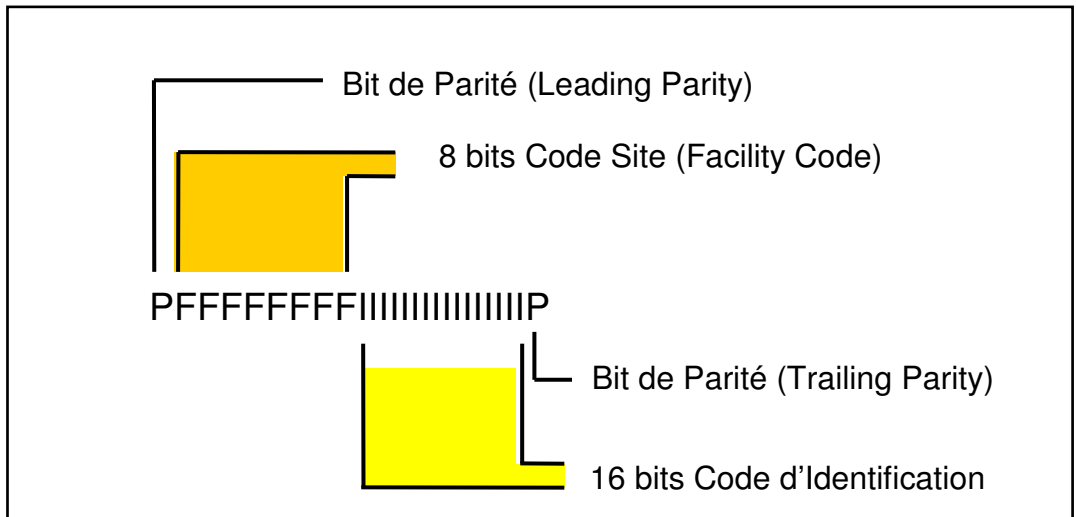

Figure 10-1: Description Format 26 Bits.

## 10.1.2  Le format 37 bits

Se donne pour objectif de garantir l'unicité des numéros. En effet, l'émission de ce type de carte est contrôlée.

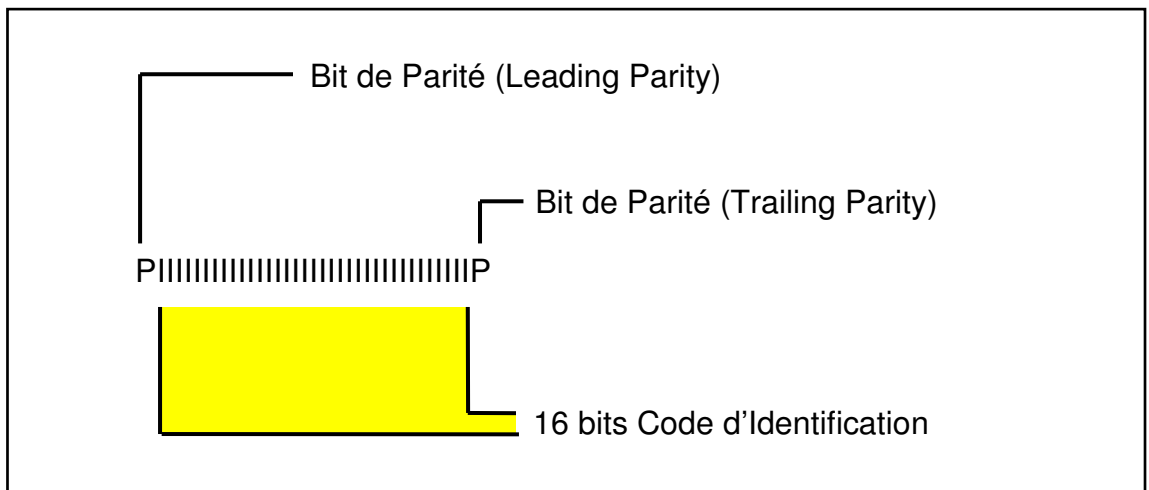

Figure 10-2: Description Format 37 Bits.

- ➢ **Bits de Parité**: utilisé pour compter le nombre des 0 et des 1 contenus dans la carte. Si ces bits sont à zéro cela signifie dans la plupart des cas qu'il y a un nombre pair de zéro ou de un dans la carte.

- **Code site** (ou Facility Code) : Initialement créé pour procurer une large production des cartes 26 bits par la duplication plutôt que l'augmentation du nombre de bits destiné à être le code de l'utilisateur. Quand un nouveau site s'installe, un code site n'ayant jamais servi dans cette zone géographique est donné pour une meilleure sécurité, ainsi tous les utilisateurs disposeront d'un seul et même site code ce qui fait que les équipements de contrôle n'auront pas à le mémoriser réduisant ainsi l'utilisation de la ressource mémoire.

- **Code d'Identification** : représente l'ID de l'employé. Il est souvent imprimé sur la carte dans le cas des cartes 26 bits.

Actuellement, le marché pour le contrôle d'accès ne cesse de croître et il devient impossible de maintenir une telle stratégie. Nombreux sont ceux qui continuent à fonctionner avec les cartes standards (26 bits) pour la simple raison que le changement reste très coûteux aux deux niveaux matériel et logiciel. Cela dit, la nouvelle tendance vise à utiliser des cartes avec un nombre de bits plus grand et à éliminer la notion de code site. Tant que le numéro de la carte est unique, la sécurité est maintenue c'est pourquoi les compagnies devraient toutes obtenir un certificat attestant que leurs cartes sont uniques et qu'elles ne seront jamais dupliquées.

## 10.2 Utilisation du SDK

Pour les besoins du développement, l'acquisition du SDK (Software Developer's Kit) s'avère utile quant à l'intégration du lecteur dans nos propres applications. Le lecteur utilise des pilotes du système d'exploitation d'où la raison de sa propriété plug and play, par contre le SDK nécessite l'utilisation d'une DLL « pcProxAPI.DLL ». Le rôle de la bibliothèque est de se connecter automatiquement au port USB, effectuer les opérations nécessaires à la lecture et écriture de la configuration du dispositif et récupérer l'identifiant de la carte à chaque fois que cette dernière est lue par le lecteur.

Le rôle du contrôle Activex inclus dans ce SDK est d'interpréter les données de la carte en filtrant les bits de parité, les bits de l'ID et les bits de Facility Code s'il y en a.

Les éléments de configuration sont regroupés en un ensemble de structures et de fonctions. Avant de pouvoir y accéder, la DLL doit effectuer la connexion via le port USB et ensuite lire l'ensemble des éléments de configurations pour pouvoir les manipuler grâce à un ensemble de fonctions « SET » et « GET »

En résumé, le SDK est un code source contenant des fonctions qui font appel à la DLL formant ainsi une couche d'abstraction. Cette dernière s'occupe de communiquer directement avec le Kit.

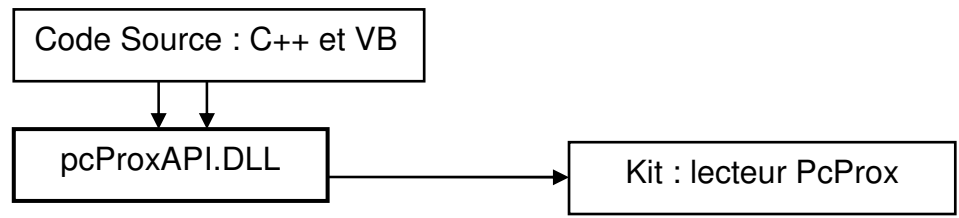

Figure 10-3: Schéma de déploiement du SDK.

L'assimilation du fonctionnement a permis de déterminer les fonctions responsables de la configuration du dispositif ainsi que la démarche suivie par le SDK pour être averti du passage de la carte et récupérer par la suite l'Identifiant. Toutes les fonctions responsables de la configuration du dispositif sont déclarées dans un module comme étant des fonctions de la bibliothèque. Essentiellement, on retrouve les fonctions suivantes:

- **SetIDDispParms/GetIDDispParms** :
    - Configure le caractère à envoyer entre les bits de l'Identifiant et les bits du Facility Code
    - Configure le caractère à envoyer à la fin de l'Identifiant
    - Configure la taille de l'Identifiant (ID)
    - Configure la taille du Facility Code (FAC)
- **SETLUID/GETLUID** :
    - Configure le nombre de dispositif en marche en cas de multi-device.
- **SETLEDCtrl**:
    - Configure le LED : rouge, vert ou par défaut

- **SETFLAGS2/GETFLAGS2**:
  - Introduit des caractères au début des données de la cartes
  - Décide si la sortie sera hexadécimale
  - Beep quand l'Identifiant est reçu
- **SetFlags3**:
  - Décide de l'affichage de l'ID en décimale ou en hexadécimale
  - Décide de l'affichage du FAC en décimale ou en hexadécimale
- **SetIDDispParms2**:
  - Les caractères de début des données (leur nombre doit être <= 3)
- **SetIDDispParms3**:
  - Les caractères de fin des données (leur nombre doit être <=3). Le nombre de caractères introduits au début + le nombre de caractères introduits à la fin des données de la carte doivent être inférieur ou égal à 3 avec une priorité aux caractères de début.
- **SetIDBitCnts**:
  - Le nombre de bits de parité situés au début des données pour être enlevé
  - Le nombre de bits de parité situé à la fin des données pour être enlevé
  - Le nombre de bits ID + FAC
  - Le nombre total de bits
- **SetFlags**:
  - Configure la taille des données à envoyer
  - Valider la lecture d'une carte selon que cette dernière satisfasse un certains nombre de bits
  - Retirer le code site (FAC) de l'ID
  - Envoyer le code site
  - Mettre un délimiteur entre le code site et le code ID

- o Utiliser un caractère de fin de ligne (entrée par défaut)
- o Choisir d'envoyer l'ID une fois reçu ou attendre
- o Choisir de ne pas envoyer les données au port USB

➢ **SetTimeParms**:
- o Configure le temps que le lecteur prendra avant de recevoir les données d'une autre carte
- o Configure le temps pendant lequel le lecteur garde les informations de la carte qui vient d'être lue.

➢ **SetActDevice**:
- o Configure lequel des lecteurs sera actif ( le cas de mutli-device)

➢ **ResetFactoryDflts**:
- o Réinitialise la configuration aux valeurs par défaut

À la fin des modifications, un appel à la fonction **writeCfg()** qui est une fonction de la librairie, doit être effectué afin que le dispositif prennent en considération la nouvelle configuration de façon permanente non volatile.

Maintenant qu'on s'est familiarisé avec les fonctions du SDK, il est devenu plus facile d'encapsuler l'essentiel de ses fonctionnalités afin d'obtenir un composant avec une interface facile à assimiler et qui permettent de configurer le lecteur et de récupérer l'ID de la carte rapidement.

Nombreuses sont les techniques qui permettent un tel résultat, entre autres les applets Java, les JavaBeans. Leur but étant de faciliter la communication entre les différents composants d'un logiciel.

## 10.3 Encapsulation:

Etant donné les avantages incontournables qu'offre cette technique, son utilisation est devenue monnaie courante quand il s'agit de développer des logiciels. En effet, elle offre une stabilité et une protection contre les erreurs (à ne pas confondre avec la sécurité, chose qu'elle n'assure aucunement). En effet, le développement par composants permet à un logiciel complexe d'être étendue et maintenue facilement.

Plusieurs études ont démontrées que le coût élevé d'un logiciel n'est pas dû à son développement mais aux heures consacrées à sa maintenance. Les composants qui ont été bien encapsulés sont de loin les plus faciles à maintenir.

Par ailleurs, étendre les fonctionnalités d'un logiciel une fois qu'il a été mis en place peut mener à un dysfonctionnement de ses autres parties. L'encapsulation permet de réduire ce risque. Son principe consiste à séparer les parties volatiles qui sont les détails de l'implémentation de la partie stable qui est l'interface limitant ainsi les dépendances entre les composants.

Le composant encapsulé en cachant son fonctionnement interne, agit comme une boite noire qui interagit avec les autres composants en leur offrant un service. Encapsuler les fonctionnalités du SDK revient à cacher celles jugées inutiles et à procurer une interface simplifiée rendant uniquement les services voulus.

L'interface du contrôle Activex permet de se connecter au lecteur et d'afficher l'Identifiant à chaque lecture de badge.

### 10.3.1 Choix du Contrôle Activex

La technologie Activex a été pensée pour permettre à différents logiciels écrits par des sociétés différentes de communiquer entre eux. Un contrôle Activex est une bibliothèque de commandes utilisables par les logiciels. Le fait de pouvoir ainsi disposer de commandes toutes faites augmente la rapidité de développement, il n'est plus besoin de réécrire des composants déjà écrits par d'autres, il suffit de les appeler en utilisant les bonnes fonctions et les bons paramètres.

Il est vrai qu'un contrôle Activex est bien plus rapide qu'un Activex EXE mais qu'il n'est pas aussi rapide qu'une DLL, il dispose néanmoins de sa propre interface utilisateur et peut interagir avec l'application le contenant par différentes façons.

Ses propriétés offrent plus de souplesse quand à la modification des valeurs en mode conception et son grand avantage réside dans son indépendance par rapport au langage avec lequel il é été développé. Destiné à être utilisé dans n'importe quel langage sous la plate-forme Windows.

Grand avantage aussi c'est la rapidité car contrairement à java, un contrôle Activex est un programme compilé. Etant un programme win32, il peut facilement effectuer des tâches administratives

### 10.3.2 Construction du Contrôle Activex : PcProxActivex

Ce contrôle a été conçu avec l'idée d'offrir à son utilisateur final (qui sera aussi un développeur) un composant qui lui permettra d'interagir avec le lecteur :

➢ **En mode Exécution**:

- o Se connecter
- o Récupérer l'Identifiant de la carte

➢ **En mode Conception**:

- o Modifier la configuration du lecteur

### 10.3.3 Les étapes de création d'un contrôle Activex

D'abord, on crée le squelette du contrôle c'est à dire les éléments de l'interface sans aucun code derrière. On obtient ainsi un aperçu de l'interface graphique du contrôle.

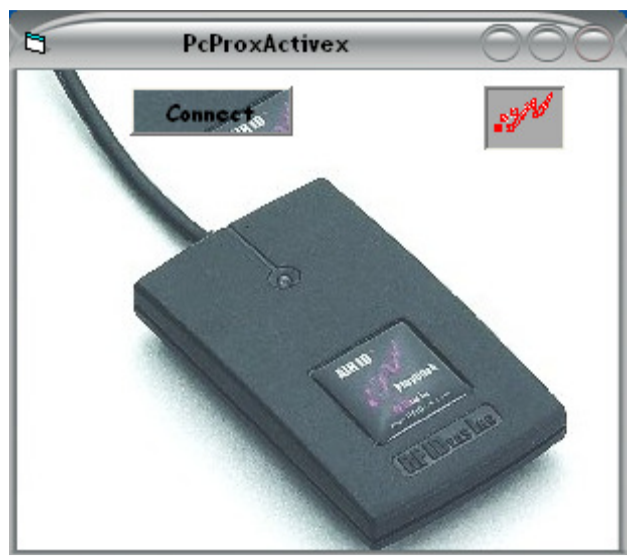

Figure 10-4: PcProx.

Ensuite, on ajoute les propriétés, les événements et les fonctions. Le contrôle PcProxActivex compte un ensemble de 6 propriétés représentant les éléments de configuration du lecteur pcProx :

### 10.3.4 Les propriétés

- **Beep :** configure le comportement du Beep en mode conception. La valeur de cette propriété est mise à True par défaut ce qui signifie que le Beep est actif. Le changement de cette propriété prend effet lors de l'exécution du contrôle.

- **LED :** configure le LED du lecteur. La valeur par défaut repésente la couleur Rouge en cas d'inactivité et vert lors du passage d'une carte. Les autres Options (Red, Green or BothColors) masque l'activité du lecteur en gardant toujours la même couleur.

- **La propriété AcceptCardData :** le lecteur est configuré pour accepter tout type de cartes, toutefois vous pouvez, grâce à cette propriété, restreindre la lecture au seul type Standard 26 bits ou bien au type Unique 37 bits.

- **La propriété ValidDataTime :** le champ de cette propriété représente la durée pendant laquelle les données de la carte sont gardées dans la mémoire du lecteur. Le lecteur ne peut donc commencer à relire une autre carte (ou bien la même carte) qu'à la fin de cette durée. C'est aussi la durée pendant laquelle le LED reste en vert lors de la lecture d'une carte. Le temps minimum qu'un lecteur peut accepter est 900 ms. Si vous entrez donc une valeur inférieure à 900 elle ne sera pas accepter et le champ prendra instantanément la valeur 900.

- **La propriété LeadingParity :** Initialement à 0. Si un chiffre n différent de 0 est mis dans ce champs, le lecteur sera informé que les premiers nième bits sont des bits de parité.

- **La propriété TrailingParity :** Initialement à 0. Si un chiffre n différent de 0 est mis dans ce champ, le lecteur sera informé que les derniers nième bits sont des bits de parité.

La combinaison résultante des valeurs initiales des propriétés constituent la configuration par défaut du lecteur et donc le déploiement du contrôle peut très bien se faire sans aucune modification au préalable. Il suffira dans ce cas de l'exécuter et de récupérer l'Identifiant instantanément à chaque lecture de carte par le lecteur PcProx.

### 10.3.5 Les événements

Le contrôle Activex génère un seul événement qui est GetID : C'est cet événement qui permet de récupérer l'Identifiant de la carte grâce au paramètre de la fonction getID.

### 10.3.6  Les méthodes

L'utilisation de ces méthodes n'est pas nécessaire au fonctionnement du contrôle Activex. La plupart ont été ajouté pour les besoins du prochain composant.

- **AddFromat :** le contrôle Activex reconnait par défaut le format 26 bits (appelé Standard) et le format 37 bits (appelé Unique). Le format 26 bits étant le plus communément utilisé. Cette fonction permet d'étendre l'utilisation du lecteur à d'autres cartes de nombres de bits différents.

En effet, il suffit d'ajouter une seule ligne : (par exemple)

MyCtrl.AddFormat(« PPFFFFFIIIIIIIIIIIIIIIIIIIIIIIIIIP »)

Ceci est un exemple de format pour une carte 34 bits.

La lettre P : représente les bits de parité

La lettre F : représente les bits de Facility Code

La lettre I : représente les bits ID

Si la saisie du format contient d'autres lettres que celles citées ci-dessus, un message d'erreur s'affiche indiquant que le format est invalide.

On peut également ajouter un format sans les bits de parité si on remplit les deux champs de

« LeadingParity » et « TrailingParity »

 Dans l'exemple précédent, on aura :

MyCtrl.AddFormat(« FFFFFIIIIIIIIIIIIIIIIIIIIIIIIIII»)

Avec le champs de LeadingParity à 2 et le Champ de TrailingParity à 1.

- **Connect :** cette fonction fait en sorte que le contrôle Activex se connecte automatiquement dès son exécution au lieu de presser le bouton « Connect », si son paramètre est à True.
    - MyCtrl.Connect(True)

- **Beep_control :** cette fonction joue le même rôle que la propriété Beep à la seule différence qu'elle permet d'effectuer les modifications en pleine exécution du contrôle. On peut à présent désactiver le Beep en décochant une case de l'interface au lieu d'être obligé d'arrêter l'exécution et de changer la valeur de la propriété en mode conception.
    - MyCtrl.Beep_control(False)

- **Masquer_ID :** certaines applications pourraient ne pas solliciter un affichage de l'Identifiant par mesure de sécurité. Cette fonction permet donc, comme son nom l'indique, de le masquer.
    - MyCtrl.Masquer_ID(True)
- **AddLeadChars :** afin de rendre l'identifiant difficile à reproduire à partir du clavier, on peut décider d'ajouter des caractères au tout début. Le nombre total des caractères à ajouter est limité à trois.
- **StripLeadChars :** après avoir ajouter les caractères de début, on peut décider de les enlever par la suite. Cette fonction permet de restaurer l'état initial de la récupération des données de la carte.

### 10.3.7 Synthèse

Le SDK est mal documenté, difficile à comprendre et nécessite un temps non négligeable pour réussir à saisir le fonctionnement et à récupérer les informations souhaitées. En l'encapsulant par le moyen d'un contrôle Activex, on réussit à obtenir un produit simple, pratique et efficace.

## 10.4 Empaquetage

Une fois le développement de notre contrôle Activex terminé, il faut engendrer le fichier .OCX qui est une sorte de bibliothèque contenant le programme du contrôle déjà compilé. A partir de ce moment on peut utiliser le contrôle indépendamment du langage avec lequel il a été écrit.

Reste un petit inconvénient, si l'on souhaite déployer notre contrôle Activex sur une autre machine, il faut obligatoirement passer par l'étape : enregistrement du contrôle dans la base de registre.

Cet enregistrement se fait automatiquement uniquement lorsque :

- Le projet se compile et crée le fichier ocx
- Le projet ouvert utilise un ocx qui se trouve dans le même répertoire
- Le programme exécuté utilise un ocx qui se trouve dans Le même répertoire

> Le programme exécuté utilise un ocx qui se trouve dans le même répertoire system de Windows

Sinon, il faudra l'effectuer manuellement en tapant la commande : Regsvr32 <nom_control> <chemin control>, ce qui nous donne le message suivant :

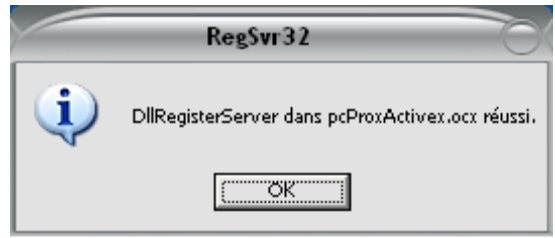

Figure 10-5: Enregistrement du composant Activex.

A la suite d'un enregistrement, il n'est plus possible de déplacer le fichier OCX. La référence contenue dans la base de registre n'étant plus valide, les programmes ne pourront plus fonctionner.

Il ne faut pas oublier que le contrôle à d'autres dépendances comme le fichier de la DLL « pcProxAPI.DLL » et le fichier ocx « RFIDFilter.ocx » qui doivent se trouver dans le même emplacement que lui et subir la même étape d'enregistrement pour ce qui est du fichier ocx, afin d'assurer son bon fonctionnement. Toutes ces contraintes n'avantage nullement le contrôle qu'on a construit et qu'on voulait simple et pratique.

Afin de palier à ce problème, une solution d'empaquetage à été mis en œuvre. En effet, l'assistant empaquetage et déploiement prend en charge l'enregistrement ainsi que les références vers les fichiers de dépendance en mettant le tout dans un package facile à déployer grâce à un fichier d'installation Setup.

### 10.4.1   Utilisation de l'assistant Empaquetage et déploiement

La première étape consiste à sélectionner le projet contenant notre ActiveX Ensuite, choisir l'option empaquetage. Et par la suite l'option empaquetage d'installation standard. Il faut ensuite sélectionner un emplacement d'enregistrement pour le package. Si le contrôle sera utilisé dans un environnement autre que Visual Basic, il faut inclure la bibliothèque de dépendance Property Page. Enfin, il faut compléter les informations sur les dépendances...

Le système offre la possibilité d'inclure des fichiers de dépendances autres que « pcProxAPI.DLL » et « RFIDproxFilter.ocx » nécessaire à l'utilisation du contrôle ActiveX dans un environnement de conception.

Par défaut, l'assistant empaquetage et déploiement installe le contrôle Activex dans son propre dossier sur l'ordinateur cible

### 10.4.2 Installation du package

Une fois cette opération terminée, on obtient un package dans l'emplacement spécifié qui contient toutes les informations nécessaires à la distribution du contrôle Activex. Il suffit à présent de mettre le package sur un ordinateur quelconque (muni d'un système d'exploitation Windows) et de démarrer l'application setup. Cette dernière renseigne clairement sur la démarche à suivre d'autant plus qu'elle ne nécessite pas plus de 30 secondes.

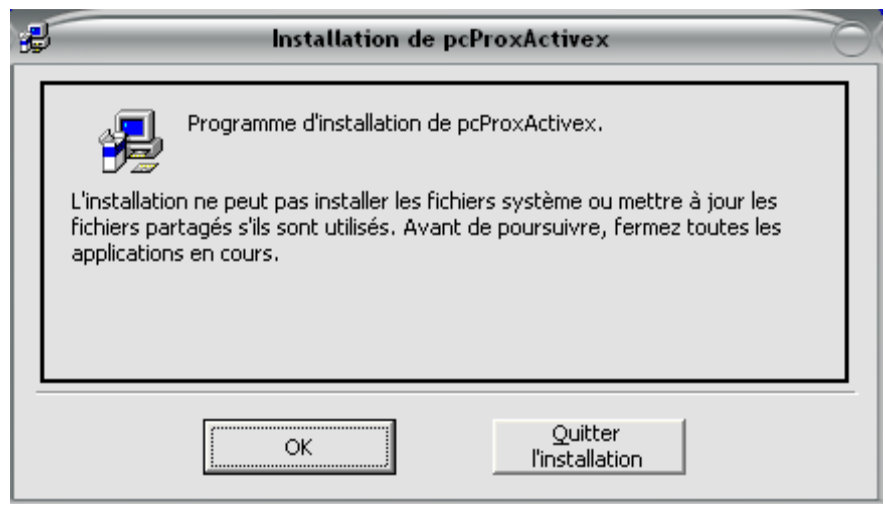

Figure 10-6: Installation du composant pcProxActivex.

Une fois qu'on clique sur ok, la fenêtre suivante apparaît. On peut choisir de modifier l'emplacement de destination ou laisser le chemin spécifié par défaut. Le démarrage de l'installation se fait en cliquant sur le bouton graphique comme indiqué ci-dessous :

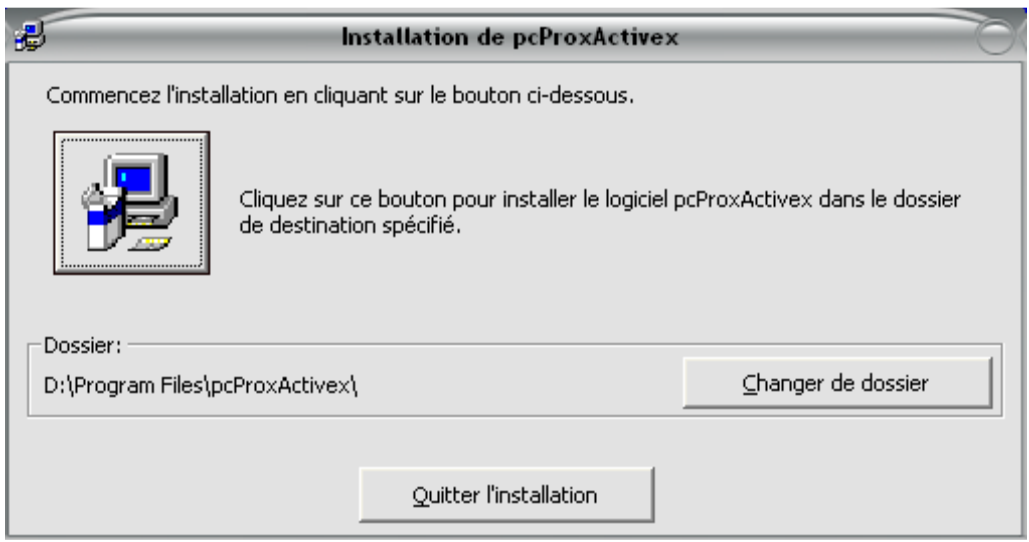

Figure 10-7: Choix de l'emplacement de l'installation.

Un bref téléchargement se fait avant d'obtenir le message suivant :

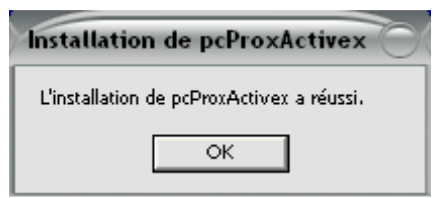

Figure 10-8: Fin de l'installation du composant.

Le contrôle Activex est maintenant enregistré dans la base de registre sans que l'utilisateur n'ait à se soucier de son enregistrement ni de connaître ses dépendances.

## 10.5 Les composants utilisés

Concrètement, on dispose de deux éléments physiques :

➢ **Lecteur RFID**: connecté à un ordinateur et dispose d'une antenne

➢ **Badge RFID**: dispose également d'une antenne en spirale connectée à une puce électronique.

La communication entre ces deux éléments n'est pas filaire et ne nécessite pas une ligne de vision entre eux contrairement à la technologie des codes barres Lors de sa connexion à l'ordinateur, le lecteur crée un champ magnétique grâce à son antenne. Ce champs transfert de l'énergie à l'antenne incorporée dans le badge RFID ce qui déclenche le « réveil » de la puce électronique. Cette dernière peut à présent communiquer son identité. Lorsque le champ magnétique cesse, l'activité de la puce s'arrête.

RFID est l'acronyme de Radio Frequency Identification. Un des membres de la famille AIDC technologies (Automatic Identification and Data Capture). Un moyen rapide et sûr d'identification. L'avantage évident de la technologie RFID par rapport à ses prédécesseurs est sa rapidité et sa facilité d'utilisation d'autant plus qu'aucun contact physique n'est requise.

On retrouve généralement deux genres d'étiquettes RFID : les étiquettes passives et les étiquettes actives.

1. Les étiquettes passives utilisent l'énergie du lecteur pour s'activer.
2. Les étiquettes actives sont munies de leur propre batterie (souvent remplaçable)

Le prix de ces derniers est par conséquent plus élevé mais leur portée est plus longue que les étiquettes passives. Les deux types de badges passifs et actifs sont utilisés dans le contrôle d'accès, à la seule différence que les premiers doivent être ramené à une distance proche du lecteur alors que les deuxièmes doivent uniquement être porté sur soi et ne nécessite aucune manipulation de la main.

Les badges dont nous disposons sont du genre passif, ils requièrent la proximité du lecteur. Ne fonctionnant avec aucune batterie, ils ont une durée de vie quasi-illimitée.

## 11 Contrôle d'Accès

La plus vieille méthode déployée dans ce domaine fût certainement la serrurerie. Le besoin de remédier à ses nombreux failles et inconvénients conduisit au développement de nouvelles technologie qui, on le constate de plus en plus dans la vie quotidienne, tendent à remplacer progressivement cette fameuse clé traditionnelle. Les cartes magnétiques et optiques voient le jour. Largement utilisée par les hôtels, elles ont l'avantage d'être neutralisées ou remplacées rapidement lors d'une perte. L'évolution de la carte fait en sorte que cette dernière passe d'un simple support magnétique ou optique à une puce informatique ne se contentant plus de simplement véhiculer un code mais de communiquer avec le système gérant les accès ou subir une nouvelle programmation afin de déjouer les copies.

Les cartes à puce intelligentes (Smart Cards) quant à elles, contrairement aux cartes magnétiques et optiques, dispose d'un microprocesseur interne qui lui permet d'effectuer des traitements sur l'information qu'elle reçoit, à noter aussi que les données stockées sur la carte sont cryptées.

Une faille évidente de l'utilisation des cartes est qu'elles n'assurent pas l'authentification. Un système de gestion des accès ne peut donc être certain que la personne en possession de la carte est bien son réel propriétaire.

A la fin du $19^{ème}$ siècle, Alphonse Bertillon, un criminologue français, découvrit qu'en prenant quatorze mensurations (taille, pieds, main, nez, oreilles, etc,...) sur n'importe quel individu, les chances sont presque nulles de retrouver les mêmes chez une autre personne. Ce fût la naissance d'une technologie de plus en plus populaire de nos jours et qui n'est autre que la biométrie, définit comme étant toute identification par mesure du corps.

Considéré comme un moyen sûr d'identification, puisque la biométrie moléculaire nous renseigne que chaque être humain sur terre possède un patrimoine génétique unique influant sur les caractéristiques morphologiques, elle présente l'énorme avantage de se passer des mots de passe ou de tout matériel physique nécessaire à l'identification.

La liste des techniques est longue, on retrouve les empreintes digitales, l'iris, les réseaux veineux de la rétine, la forme de la main, les traits du visage, la reconnaissance de la voix... les moyens les plus utilisés restent cependant les empreintes digitales et les empreintes rétiniennes.

## 11.1 Description du composant Lock/Unlock

Il est connu que le choix d'un mot de passe ne doit surtout pas être des noms, des dates ou des mots du dictionnaire. Pour assurer sa robustesse, il faudrait inclure des lettres en majuscule et en minuscule, des chiffres et au moins un caractère spécial. Seulement, la complexité d'un tel mot de passe se heurte à la difficulté de sa mémorisation et surtout le désagrément que cela inflige de devoir le saisir à chaque fois. La plupart des gens utilisent par conséquent des mots de passe tellement simple qu'ils peuvent être piratés en un temps record.

D'où le grand avantage que l'on puisse tirer de l'utilisation des badges qui en plus de remplacer le mot de passe peuvent introduire des caractères spéciaux difficiles à reproduire à partir du clavier.

Sans oublier que nombreux sont ceux qui oublient de fermer leur session et laisse par conséquent leurs données sans aucune protection ce qui peut s'avérer dans certains environnement très risqué même pour une durée très minime.

Ce n'est certainement pas en identifiant un badge qu'on pourra dire qu'on a identifié la personne. D'autant plus qu'il n'est pas improbable qu'un identifiant transmis par la puce puisse être intercepté afin d'être cloné pour de l'usurpation d'identité. C'est pour cette raison qu'il est préférable d'augmenter la sécurité en combinant l'identification par carte et par mot de passe. Ainsi, une personne pour s'authentifier aura besoin de présenter son badge (un matériel qu'elle doit être la seule à posséder) et entrer un mot de passe (une information qu'elle doit être la seule à connaître). Le concept est certes très proche de celui de la carte bancaire où on retrouve la responsabilisation du propriétaire tout en assurant la sécurité en cas de perte ou de vol.

### 11.1.1 Utilisation du composant précédent

La personne souhaitant verrouiller et déverrouiller la machine avec son badge, doit d'abord être identifiée comme étant la personne en droit d'effectuer cette opération. D'où l'utilité d'intégrer notre composant précédent. En effet, le contrôle Activex va permettre à l'application de récupérer instantanément l'Identifiant de la carte.

La plupart des mesures de sécurité consiste à authentifier les personnes et contrôler leur droit d'accès, mais toutes ces mesures ne servent strictement à rien si l'utilisateur légitime laisse sa session ouverte même pour un lapse de temps relativement court.

Après avoir assuré l'identification grâce au premier composant, on vise à présent à atteindre l'authentification qui va nous permettre de vérifier l'identité présumée des utilisateurs.

Selon l'environnement dans lequel les utilisateurs se trouvent, ils auront à la fois besoin d'une protection simple ou une protection forte.

L'Authentification simple consiste à utiliser uniquement le badge ID pour garantir l'accès à la machine, utile quand il s'agit d'absences de durées très courtes ou de simples déplacements qui implique un éloignement de l'ordinateur.

Par contre l'Authentification forte réside dans l'utilisation de deux facteurs d'identification qui sont et le badge ID et le mot de passe.

Différents sont les moyens offerts par le système d'exploitation Windows afin de verrouiller la session d'un utilisateur.

> ➢ **Par clavier :**

Le moyen le plus simple et le plus rapide puisqu'il suffit de presser la touche du logo Windows suivi de la touche L.

> ➢ **Par raccourci :**

Dans le cas d'un clavier ne possédant pas de touche Windows ou simplement pour l'envie d'utiliser une autre méthode que celle du clavier, on peut toujours créer un raccourci sur le bureau.

Néanmoins toutes ces méthodes font recours à une seule et même ligne de commande :

**rundll32.exe user32.dll, LockWorkStation**

Dans un but d'efficacité et d'automatisation, l'application du contrôle d'accès basé sur la proximité cherche avant tout à rendre l'opération de verrouillage/déverrouillage de l'ordinateur à la fois simple et agréable tout en garantissant une forte sécurité.

Lors de la proximité de la carte du lecteur PcProx, le champ relatif au mot de passe Windows est rempli instantanément par l'identifiant de la carte, le lecteur s'occupe également d'envoyer le caractère « Entrée » à la fin pour que l'utilisateur accède à sa session sans aucune intervention manuelle autre que la présentation du badge au lecteur. D'autre part, comme l'application présente une option supplémentaire visant à renforcer le contrôle d'accès, l'utilisateur aura aussi la possibilité d'ajouter un mot de passe ce qui implique que lors du déverrouillage, le mot de passe connu uniquement par l'utilisateur devra être saisi après avoir présenté le badge au lecteur.

### 11.1.2  Simulation de l'éloignement physique de la carte

Dans l'absence du sonar censé détecter l'éloignement de la carte et par conséquent de son porteur afin d'effectuer le verrouillage de la machine, on présente le badge pour une deuxième lecture afin de simuler la détection de l'éloignement de la personne. Ainsi, quand la session est ouverte et que le lecteur reçoit l'identifiant de l'utilisateur courant, le verrouillage a lieu instantanément.

## 11.2 Sécurité et Confidentialité

### 11.2.1 Menaces et Risques

Le nombre de personnes ayant été victimes du vol de leur ordinateur portable ne cesse de croître alors que les chances de le récupérer restent quant à elles très minimes.

Les conséquences sont d'autant plus désastreuses quand il s'agit d'un ordinateur qui contient des informations personnelles de valeurs pouvant créer des dommages si elles tombaient dans de mauvaises mains. L'une des conséquences la plus courante est l'usurpation d'identité.

Une protection physique de l'ordinateur est hautement recommandée, cela dit il est toujours utile de prévoir le pire et donc d'assurer également la sécurité logique des données afin d'empêcher même en cas de vol l'accès aux données de notre ordinateur classées confidentielles.

Les pratiques communes de sécurité suggèrent plusieurs lignes de défense ou ce que l'on appelle « la défense en profondeur», un terme emprunté à une technique militaire qui vise à retarder l'ennemi. Le principe consiste en l'utilisation de plusieurs couches de protection chacune de type différent, afin de réduire le risque lorsqu'un composant de la sécurité est compromis ou défaillant.

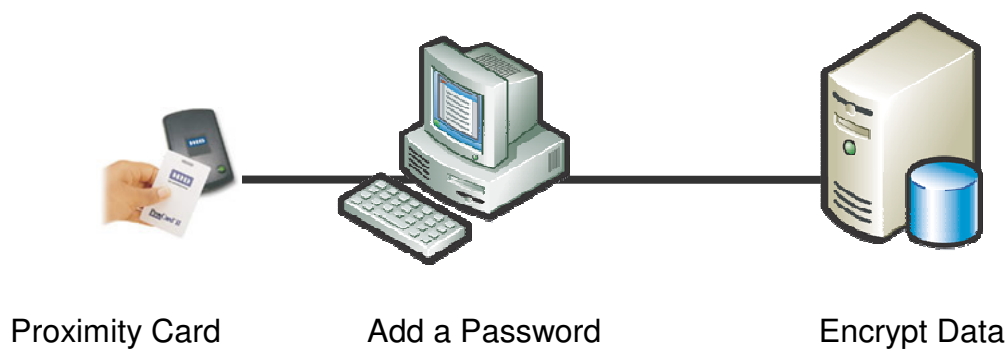

Proximity Card          Add a Password          Encrypt Data

Figure 11-1: RFID Architecture.

Le système d'exploitation Windows offre un outil de cryptage qui est l'EFS (Encryption File System). Reconnu comme étant un outil assez solide, il est cependant facilement contourné si l'on considère le fait que ce système ne fonctionne que sur une partition NTFS et donc le fait de copier les données sur une clé USB formatée en FAT donnera pleinement accès aux données puisque ces derniers seront copiés sans le cryptage.

D'autant plus, qu'en cas de vol, si l'on réussit à franchir la première barrière qui est l'accès à la session, alors ce système de cryptage ne sert strictement à rien puisque toutes les données seront visibles une fois la session ouverte.

### 11.2.2 Le Composant de Cryptage et de Décryptage

Contre toutes les menaces, le meilleur moyen d'empêcher quelqu'un de voler des données reste celui de les crypter. D'ailleurs, toutes les grandes entreprises conscientes des conséquences dues à des fuites se mettent à crypter ses données sensibles systématiquement en complément des outils de protection dont elles disposent. Ainsi, ils deviennent non utilisables sans la connaissance de la clé.

Ici le composant de Cryptage/Décryptage se base sur l'algorithme rc4 (Rivest Cypher 4), mis au point par Ronald Rivest, l'un des inventeurs du RSA, en 1987. Cet algorithme utilise normalement des clefs de taille 64 bits ou 128 bits.

De par son appartenance à la cryptographie symétrique (autrement dit à clé secrète ou privée), RC4 se sert d'une seule et même clef pour le chiffrement et le déchiffrement.

N'étant pas un algorithme très robuste, il est néanmoins utilisé dans différentes normes telles que WEP pour le 802.1 réseau Wi-Fi et le SSL et dans certaines applications commerciales telles que Oracle Secure SQL ou encore dans RSA SecurePC.

Son avantage réside clairement dans sa rapidité et sa facilité d'implémentation.

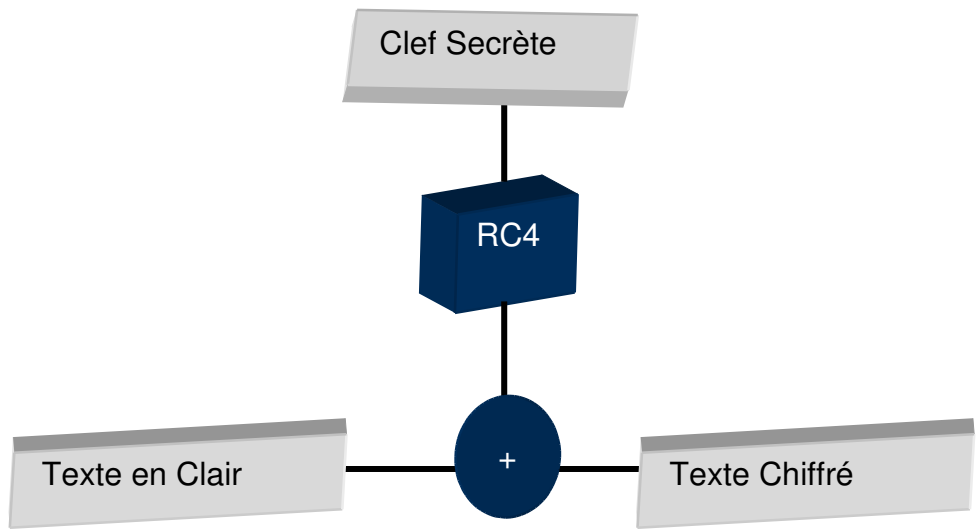

Figure 11-2: Protocole RC4.

On distingue entre deux parties principales :

- **KSA:** Key Scheduling Algorithm ou préparation de la clé
- **PRGA:** Pseudo-Random Generation Algotihm

### 11.2.2.1    Les étapes

- Initialiser un tableau de 256 bits
- Effectuer la partie KSA
- Effectuer la partie PRGA sur la sortie du KSA afin de générer la clef
- Effectuer un XOR entre les données en clair et la clef obtenue à l'étape précédente

### 11.2.2.2    Initialisation:

La clé sert à initialiser une "table d'états" de taille 256 bits, en répétant la clé autant de fois que nécessaire pour remplir la table.

### 11.2.2.3    KSA

Utilise la clef secrète pour modifier l'ordre des éléments du tableau S

```
pour i de 0 à 255
    S[i] := i
finpour
j := 0
pour i de 0 à 255
    j := (j + S[i] + key[i mod key_lentgh]) mod 256
    swap(S[i],S[j])
finpour
```

### 11.2.2.4    PRGA

Après avoir subi l'étape KSA, le tableau S maintenant mélangé est utilisé pour générer la clef qui sera additionnée au texte clair par un XOR.

```
i := 0
j := 0
tant_que générer une sortie:
    i := (i + 1) mod 256
    j := (j + S[i]) mod 256
    swap(S[i],S[j])
    octet_chiffrement = S[(S[i] + S[j]) mod 256]
    result_chiffré = octet_chiffrement XOR octet_message
fintant_que
```

Cet algorithme garantit que chaque valeur de $S$ est échangée au moins une fois toutes les 256 itérations.

## 11.2.3  Fonction du composant

Lors du cryptage, trois opérations ont lieu :

➢ La sauvegarde de la structure des répertoires à crypter.

➢ Chiffrement des données

➢ Création d'un fichier où seront sauvegardé la structure et les données cryptées ainsi que la suppression des répertoires originaux.

Lors du décryptage, les opérations précédentes prennent le sens inverse :

➢ Restauration de la structure des répertoires à partir du fichier sauvegardé

- ➢ Déchiffrement des données
- ➢ Restauration des données selon leur structure originale et suppression du fichier sauvegardé lors du chiffrement.

Concrètement l'utilisateur quand il coche l'option de chiffrement, ses données disparaissent. Elles sont par conséquent remplacer par un symbole indiquant leur état crypté. Quand l'utilisateur souhaite déchiffrer ses données, le symbole précédent disparaît pour laisser la place aux répertoires restaurés et décryptés tels qu'ils étaient avant le chiffrement.

Le programme comporte les fonctions suivantes :

- ➢ **Rc4_Crypt:** contient l'algorithme de chiffrement/déchiffrement d'un flux de données. Elle a trois paramètres qui sont le « buffer »( un espace mémoire alloué dynamiquement), la clef avec laquelle le chiffrement sera effectué et la taille des données à chiffrer ce qui revient à dire la taille du « buffer ».

  void rc4_Crypt(unsigned char* buffer, unsigned char* key, unsigned int * size)

- ➢ **saveDirectory:** la fonction qui permet de sauvegarder le nom du répertoire, son niveau dans la hiérarchie ainsi que son type qui sera par définition D pour indiquer que c'est un dossier (Directory).

  void saveDirectory(char* fileName,int depth,char type)

- ➢ **saveFile :** la fonction qui permet de sauvegarder le nom du fichier, son niveau dans la hiérarchie, son type qui sera dans ce cas F pour indiquer que c'est un fichier et enfin sa taille.

  void saveFile(char* fileName,int depth,char type,int sizeFile)

- ➢ **saveDirectoryStructure :** la fonction qui sauvegarde la structure des répertoires ainsi que des fichiers que chaque répertoire contient. Les informations obtenues constituent l'en-tête du fichier sauvegardé après chiffrement. Ce dernier permet de restaurer les données exactement de la manière avec laquelle elles étaient organisées au départ.

  void saveDirectoryStructure(char *fileName,int depth)

- **Crypt_File** : ayant déjà implémenté l'algorithme de cryptage rc4 sur un flux de données. On se contente ici de l'appliquer au flux récupéré suite à la lecture d'un fichier.

    void Crypt_File(FILE *fichier,unsigned char * key)

- **remove_directory** : cette fonction s'occupe de la suppression du répertoire racine ainsi que tout son contenu de façon irréversible de la machine afin de ne laisser aucune trace des données en clair.

    int remove_directory(char const *name)

- **createStructureFile** : la fonction qui crée le fichier nommé « Structure ». ce dernier contient une partie en-tête qui renseigne sur la hiérarchie des répertoires, des fichiers contenus dans chacun des répertoires ainsi que sur la taille de chaque fichier. Ensuite vient la partie de données cryptées.

    void createStructureFile()

- **restoreDirectoryStructure** : utilise l'en-tête du fichier sauvegardé après chiffrement pour retrouver la structure originale des répertoires et des fichiers

    void restoreDirectoryStructure(unsigned char *key)

Toutes ces fonctions citées précédemment sont regroupées dans une bibliothèque nommée « arcfour.dll ». Cette dernière dispose de trois fonctions supplémentaires qui utilisent les autres fonctions et sont les seuls à être exportées :

```
void CryptData(unsigned char* key);
void CryptFileStructure(unsigned char* key);
void RestoreData(unsigned char* key);
```

L'algorithme étant un algorithme à clé secrète, la clef pour crypter doit être la même que pour décrypter. Par conséquent, le paramètre key de la fonction RestoreData doit correspondre au paramètre key de la fonction CryptData.

Pour crypter le fichier contenant les informations sur la structure des répertoires, on utilisera la fonction CryptFileStructure(), la clef en paramètre peut être différente que celle avec laquelle on a crypté et décrypté les données.

Pour décrypter ce même fichier, on fait appel à la même fonction avec la même clé comme paramètre.

### 11.2.4 La démarche à suivre

#### 11.2.4.1    Cryptage:

D'abord on crypte les données grâce à la fonction CryptData, on obtient un fichier nommé « Structure ». Ce dernier contient deux parties :

- Une partie En-tête: où est sauvegardé la structure des répertoires et des fichiers ainsi que la taille des données contenus dans chaque fichier.
- Une partie Données : où est sauvegardé le condensé des données en version cryptée.

On Crypte le fichier Structure avec la fonction CryptFileStructure, le fichier est supprimé et remplacé par un fichier du même nom avec un contenu crypté.

#### 11.2.4.2    Décryptage :

On décrypte le fichier Structure avec la fonction CryptFileStructure, en lui donnant comme paramètre la même clef utilisée pour le cryptage du fichier.

On restore la structure de nos répertoires et fichiers grâce à la fonction RestoreData, toujours en utilisant la même clef que pour le cryptage.

#### 11.2.4.3    L'appel de la dll depuis vb6 :

La bibliothèque a été écrite de manière à être compatible avec le langage vb. On pourra donc déclarer nos trois fonctions dans un module comme ci-dessous :

> Public Declare Sub CryptFileStructure Lib "Encrypt.dll" (ByRef key As String)
>
> Public Declare Sub CryptData Lib "Encrypt.dll" (ByRef key As String)
>
> Public Declare Sub RestoreData Lib "Encrypt.dll" (ByRef key As String)

**P.S :** Les paramètres ont été déclaré ByRef pour la simple raison que les fonctions en C attendent un pointeur.

#### 11.2.4.4 Emplacement de la DLL

Lors de l'exécution du programme faisant appel à la dll, Windows cherche la bibliothèque dans le répertoire courant, dans celui de l'exécutable ainsi que dans les répertoires /Windows et /Windows/System.

Une copie de la dll dans l'un de ces deux derniers répertoires peut éviter des problèmes lors du déplacement de l'application.

## 12 Conclusion et Perspectives

Le suivi quotidien de l'état du patient lui permet d'agir librement au niveau de ses actes. Chose qui n'est pas toujours autorisée pour des gens très âgés par exemple. D'où le grand intérêt de développer une application qui surveille les mouvements des patients pour une détection efficace et rapide des chutes.

Cette application permet non seulement le suivi de l'état du patient et de ses mouvements mais elle vérifie aussi si les données reçues viennent de la bonne personne, en utilisant un système d'identification par électrocardiogramme (ECG). Ainsi, le non répudiation des données est parfaitement garanti grâce au système de contrôle d'accès (RFID).

D'autre part l'implémentation du module « Communication via web » afin d'envoyer les informations aux capteurs aura une grande importance dans les travaux futurs.

## 13 Bibliographie